\documentclass[12pt,noshowpacs,nofootinbib,notitlepage,amsmath]{revtex4-1}
\usepackage{setspace}
\allowdisplaybreaks
\usepackage{graphicx,color}
\usepackage[charter]{mathdesign}
\usepackage{multirow}
\usepackage{enumerate}

\DeclareSymbolFontAlphabet{\mathcal}{symbols}
\DeclareSymbolFont{symbols}{OMS}{xmdcmsy}{m}{n}
\DeclareSymbolFont{largesymbols}{OMX}{xmdcmex}{m}{n}
\SetSymbolFont{symbols}{bold}{OMS}{xmdcmsy}{b}{n}
\def\sl#1{#1\!\!\!\slash}

\def\Mp{m_{\mathrm{Pl}}}

\def\LQG{\Lambda_{\textrm{QQG}}}
\def\sl{\ell}
\def\Mq{{\cal M}}

\begin{document}
\title{\color{blue}\Large Not quite a black hole}
\author{Bob Holdom}
\email{bob.holdom@utoronto.ca}
\author{Jing Ren}
\email{jren@physics.utoronto.ca}
\affiliation{Department of Physics, University of Toronto, Toronto, Ontario, Canada  M5S1A7}
\begin{abstract}
Astrophysical black hole candidates, although long thought to have a horizon, could be horizonless ultra-compact objects. This intriguing possibility is motivated by the black hole information paradox and a plausible fundamental connection with quantum gravity. Asymptotically free quadratic gravity is considered here as the UV completion of general relativity. A classical theory that captures its main features is used to search for solutions as sourced by matter. We find that sufficiently dense matter produces a novel horizonless configuration, the 2-2-hole, which closely matches the exterior Schwarzschild solution down to about a Planck proper length of the would-be horizon. The 2-2-hole is characterized by an interior with a shrinking volume and a seemingly innocuous timelike curvature singularity. The interior also has a novel scaling behavior with respect to the physical mass of the 2-2-hole. This leads to an extremely deep gravitational potential in which particles get efficiently trapped via collisions. As a generic static solution, the 2-2-hole may then be the nearly black endpoint of gravitational collapse. There is a considerable time delay for external probes of the 2-2-hole interior, and this determines the spacing of echoes in a post-merger gravitational wave signal. 
\end{abstract}
\maketitle

\section{Introduction and Summary}
\label{intro}

The black hole information paradox is a significant driver of current research, but any consensus as to its resolution or physical meaning has yet to be attained. One question is whether this problem is fundamental in the sense that it requires some new input from the ultraviolet (UV) completion of general relativity (GR). This would appear puzzling given the small curvature near the horizon of a large black hole. Nevertheless the interplay between string theory in particular and black hole physics has given many avenues of approach to the black hole information paradox. The fuzzball proposal is one striking example of how the physics of quantum gravity could enter to drastically modify the very meaning of space-time right at the location of the would-be horizon \cite{Mathur:2005zp}. More general arguments from effective field theory also indicate that something drastic must happen close to the horizon (e.g.~a firewall) if information is not to be lost \cite{Almheiri:2012rt,Giddings:2014ova}.

In this paper we would like to describe another example where a proposed UV completion of gravity leads directly to a drastic modification of the Schwarzschild (Schd) spacetime starting just at the location of the would-be horizon. In this case the metric description still holds in the interior large curvature region where the volume of spacetime shrinks drastically. For this gravity theory we have found what may be the generic endpoint of gravitational collapse for a general matter distribution. The vacuum Schd solution still exists, and possibly other solutions with a horizon, but these may not be the physically relevant, sourced-by-matter solutions.

It is the link between the absence of a horizon after gravitational collapse and the absence of the black hole information paradox that provides the impetus for this work. The proposed theory of quantum gravity that underlies our study is far from new, and coming along with it is a well-worn problem that needs to be faced. It could be said that we are trading one problem for another, but we would like to argue that these two problems are not equal in their intractability.

Quantum quadratic gravity (QQG) is characterized by two dimensionless couplings and one mass scale,
\begin{eqnarray}\label{eq:quadratic}
S_{\mathrm{QQG}}=\int d^4x\,\sqrt{-g}\left(\frac{1}{2}\Mq^2R-\frac{1}{2 f_2^2}C_{\mu\nu\alpha\beta}C^{\mu\nu\alpha\beta}+\frac{1}{3 f_0^2}R^2\right).
\end{eqnarray}
This action was found to be perturbatively renormalizable and  asymptotically free decades ago \cite{Stelle:1976gc, Voronov:1984kq, Fradkin:1981iu, Avramidi:1985ki}. In the standard picture the running couplings remain weak at the mass scale $\sim|f_i\Mq|$, below which the effective description is GR with $\Mq$ identified with the reduced Planck mass.  Unfortunately such a view suffers from the problem of a spin-2 ghost. Due to the higher derivative terms the propagator for the metric perturbation  on a flat background has a massive pole with negative residue in the spin-2 sector. It implies either problems with the probability interpretation and unitarity, or vacuum instability. A consensus on how to deal with this problem is still lacking.

It appears to us that the apparent intractability of this problem is linked to the assumption of weak couplings. Recently some thought has been given as to what happens if the theory enters a strong phase \cite{Holdom:2015kbf}\cite{Donoghue:2016vck}. In \cite{Holdom:2015kbf} we discussed the case where $\Mq$ is sufficiently small, so that the couplings $f_i$ grow strong and another mass scale $\LQG>\Mq$ appears. The poles in the perturbative propagator then fall into the nonperturbative region in which case the physical spectrum need not be the perturbative spectrum. A similar phenomenon occurs in quantum chromodynamics (QCD). We discussed the analogy, both the similarities and differences, between the nonperturbative graviton propagator and the nonperturbative gluon propagator. The analogy led to our conjecture that when $\Mq\lesssim \LQG$, the naive spin-2 ghost is removed and a mass gap forms as determined by $\Mq$. Since diffeomorphism invariance (like gauge invariance in QCD) is preserved we further argued that GR emerges in the infrared (IR) in the limit of a vanishing mass gap, $\Mq\to0$. (In \cite{Donoghue:2016vck} the analogy between quadratic gravity and a confining gauge theory is pursued in a somewhat different way to also arrive at the emergence of GR.) The Planck mass $\Mp$ is then associated with $\LQG$. $\Mq=0$ fits into the view that there should be no mass parameters in the fundamental action, with all mass scales, including the Planck mass, arising dynamically.

In this picture QQG approaches weak limits in both the UV and IR. In the IR the Einstein term is emerging as the leading term of an effective action that is analogous to the chiral Lagrangian in QCD. In the UV and in particular at super-Planckian curvature the asymptotically free $\Mq=0$ quadratic action (\ref{eq:quadratic}) should give a good description with small quantum corrections. This leads us to our study in this paper of a purely classical action that has the same limits. Namely it interpolates between GR and the quadratic gravity description in regimes of low and high curvature respectively. The action of classical quadratic gravity (CQG) is
\begin{eqnarray}\label{eq:QGC}
S_{\mathrm{CQG}}=\frac{1}{16\pi}\int d^4x\,\sqrt{-g}\left(\Mp^2R-\alpha C_{\mu\nu\alpha\beta}C^{\mu\nu\alpha\beta}+\beta R^2\right).
\end{eqnarray}
Our theoretical perspective justifies not having to consider higher powers of curvature in this classical action when describing solutions with arbitrarily large curvatures.

We shall be able to obtain a more general understanding of the solutions of CQG, in particular those that are sourced by matter. These solutions can be both macroscopically large and highly curved. As we have suggested these solutions should be useful approximations to the corresponding states in QQG. But they will differ in at least two respects. One is that the full effects of quantum gravity should become apparent when the curvatures are of order the Planck scale. Such curvatures will occur in our classical solutions in a shell with a relatively small thickness. We consider our classical solutions as simply a way to interpolate through such a region. The other difference is that the classical theory has Planck mass ghost instabilities, both around flat backgrounds as we have mentioned and probably also around the high curvature backgrounds of interest here. Our working assumption is that these particular instabilities are a defect of CQG, and that they do not afflict the corresponding configurations of QQG. In other words we are assuming that the QQG does support some stable macroscopically large objects that serve as the endpoint of gravitational collapse and that CQG provides a window onto the main properties of such objects.

The present knowledge of the space of static spherically symmetric solutions of CQG is still incomplete. The exact vacuum solutions of GR are present, but other solutions are not known analytically. The categorization of the possible leading terms in the series expansions of the solutions around $r=0$ was found long ago \cite{Stelle:1977ry}. The $(0,0)$ family describes those solutions that are nonsingular at the origin and it has two free parameters. The $(2,2)$ family has no analog in GR and it is characterized by a metric that is vanishing at the origin. It turns out to have five free parameters \cite{Holdom:2002xy} and in this sense it describes the most generic solutions. Asymptotically-flat solutions will quickly approach the Schd solution at large $r$, where the deviations due to the higher derivative terms in CQG are becoming exponentially small. As seen in the linearized theory such corrections are generally present when the solution is sourced by matter \cite{Stelle:1977ry}, and so the exact Schd solution does not play the same fundamental role that it plays in GR. The important question is then what solution is actually chosen for a given matter distribution.

\begin{figure}[!h]
  \centering%
{ \includegraphics[width=15cm]{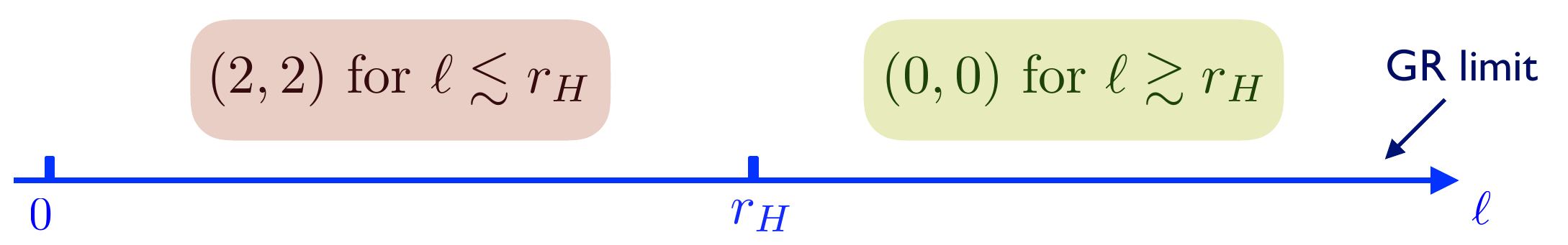}}
\caption{\label{fig:lMplot} 
A schematic illustration of asymptotically-flat horizonless solutions that couple to a thin-shell with physical mass $M$ and shell-radius $\sl$. $r_H=2M/\Mp^2$ denotes the would-be horizon.}
\end{figure}

Numerical analysis is necessary to search for solutions in the fully nonlinear theory.  To help make this problem tractable we focus on solutions induced by a thin spherical shell of matter. The answer to the above question for a thin-shell is then as illustrated in Fig.~\ref{fig:lMplot}.  When $\sl\gtrsim r_H\equiv 2M/\Mp^2$ there are the regular $(0,0)$ solutions, which can differ quite substantially from the corresponding GR solutions for $\sl$ close to $r_H$. When $\sl\lesssim r_H$ the $(2,2)$ solutions take over. These latter solutions remain horizonless, but they resemble the exterior Schd solution so well that the deviation is only visible outside the would-be horizon at a proper distance of the order of the Planck length. This feature is also expected to apply for a more general smooth matter distribution. So extremely high compactness is naturally achieved in CQG as dictated by the dynamics of gravity. In contrast the horizonless black hole mimickers constructed in GR usually rely on some fine-tuned properties in the matter sector \cite{BHM1, BHM2, BHM3, BHM4,Visser:2003ge, Visser:2009pw}. We call these new horizonless objects ``2-2-holes''. 

In the interior of a 2-2-hole, $r\lesssim r_H$, the curvatures grow large. As one of our main findings, the interior of large 2-2-holes reveals a novel scaling behavior with respect to their size or $M$, in the following sense. For the thin-shell located at a fixed $\sl/M$, curvature invariants are universal functions of $r/M$ that are independent of $M$. This is unusual when compared to the Schd metric, where nonzero curvature invariants of dimension $2n$ fall as $1/M^{2n}$ at some fixed $r/M$. The very different interior and exterior behaviors correspond to the dominance of the quadratic and linear curvature terms in the CQG action respectively. The 2-2-hole then has a nontrivial transition region connecting these different behaviors. This region and how it depends on $M$ must be explored numerically. But for the interior region we have uncovered a new spacetime that applies universally to 2-2-holes of any size. Since it involves super-Planckian curvatures it makes sense only in our context of a UV complete and asymptotically free theory of quantum gravity.

The 2-2-hole interiors are characterized by a gravitational potential that deepens for increasing $M$, and which is thus extremely deep for $M$ of astrophysical size. For instance particles falling in and colliding in the interior can yield super-Planckian center of mass energies for generic kinematics. At the same time the deep potential gives rise to a very efficient trapping mechanism. Through collisions, particles will populate a phase space such that all but a tiny fraction of the particles are trapped by an angular momentum barrier.

Compared to the Schd singularity, the center of the 2-2-hole is characterized by a less singular $C^{\mu\nu\alpha\beta}C_{\mu\nu\alpha\beta}$ and in fact the CQG action remains finite. The dominant tidal forces for finite size objects are more volume squeezing rather than shape changing, due to a more singular $R^{\mu\nu}R_{\mu\nu}$. 

The existence of a timelike singularity in the 2-2-hole indicates that the spacetime is geodesically incomplete. But it is unclear whether this problem with the motion of classical point particles is physically significant. We can probe the singularity with finite energy wave packets in a relativistic classical field theory. For example consider the case of a flat spacetime where there is a timelike singularity because a single spatial point is removed. Wave packets of finite energy propagating in this spacetime would not be affected. Within an existing mathematical formulation \cite{Wald:1980jn}, we find that the wave equation around the 2-2-hole singularity shows similar behavior and finite energy waves still have unitary evolution. This singularity does not introduce an ambiguity but it is physically detectable. In fact the resulting boundary condition at the origin leads to a discrete set of localized field modes. An entropy is associated with a thermal distribution of these modes. We find that due to the large $M$ scaling behavior of the interior of a 2-2-hole, this entropy obeys an area law when the temperature is proportional to the Hawking temperature.

The direct detection of gravitational waves by Advanced LIGO opens up the era of gravitational wave astronomy \cite{Abbott:2016blz}. Geometries around the post-merger ultra-compact object are going to be examined more closely. It was argued recently that for a horizonless ultra-compact object there is some time delay before the inner structure of the object can modify the standard black hole ringdown waveform \cite{Cardoso:2016rao, Cardoso:2016oxy}. For the 2-2-hole the extremely high compactness naturally indicates a relatively long time delay. This is the time spacing between echoes of the initial ringdown, and an initial attempt to look for echoes with a similar time delay has already been made \cite{Abedi:2016hgu}. Future data should clarify matters. In this regard it will also be important to develop a similarly complete picture for rotating 2-2-holes. Such a picture would address the question of an ergoregion, and here again the special nature of the 2-2-hole interior has interesting implications. 

The rest of the paper is organized as follows. We start Sec.~\ref{sec:SSsolution} with a brief review of the basics of static spherically symmetric solutions in CQG. We then describe thin-shell models and explain our numerical strategy to search for asymptotically-flat solutions in Sec.~\ref{sec:thin-shell}. Numerical results in the fully nonlinear theory are presented in Sec.~\ref{sec:numsolution}, which shows the complementarity between the 2-2-hole and the ordinary star as illustrated in Fig.~\ref{fig:lMplot}. The novel scaling behavior for the interior of large 2-2-holes is presented in Sec.~\ref{sec:scaling}. In Sec.~\ref{sec:phyProp} we consider various physical properties and implications of 2-2-holes.  We explore the radial stability of the 2-2-holes against movement of the shell in Sec.~\ref{sec:radialSta}.  In Sec.~\ref{sec:geodesics} we study point particle geodesics and collisions inside the 2-2-hole which leads to a discussion of the trapping mechanism.  We diagnose the timelike singularity with a focus on classical field dynamics in Sec.~\ref{sec:singularity}.  In Sec.~\ref{sec:brick} we adapt the brick wall model for black hole entropy to a discussion of entropy for 2-2-holes.  In Sec.~\ref{sec:QNMs} we estimate the size dependent time delay for 2-2-holes and further describe the wave equation. A sketch of possible behaviors of a rotating 2-2-hole is given in Sec.~\ref{sec:rotation}.

\section{Static spherically symmetric solutions}
\label{sec:SSsolution}

The general line element for a static, spherically symmetric spacetime is 
\begin{eqnarray}\label{ds2}
ds^2=-B(r)dt^2+A(r)dr^2+r^2d\theta^2+r^2\sin^2\theta d\phi^2
.\end{eqnarray}
Due to the Bianchi identity only two field equations of the action (\ref{eq:QGC}) are independent. We use the $tt$ and $rr$ components of the field equations, which can also be obtained by varying the action with respect to $B(r)$ and $A(r)$. $B(r)$ is affected by a rescaling of $t$, and this is reflected in field equations that depend only on the normalized derivatives $B^{(i)}(r)/B(r)$. By convention $B(r)$ is set to unity at infinity in asymptotically-flat solutions. For generic CQG, $\alpha\neq0, \beta\neq 0$, five initial conditions are needed to determine a solution to the field equations, namely $A''$, $A'$, $A$, $B''/B$, $B'/B$ at some value of $r$ \cite{Lu:2015psa}. We shall also consider the special case with $\beta=0$; in this case there are three initial conditions, $A'$, $A$, $B'/B$ \cite{Lu:2015psa}. 

The solutions can be classified by the series expansion around $r = 0$ \cite{Stelle:1977ry}. 
\begin{eqnarray}
A(r)&=&a_s r^s+a_{s+1}r^{s+1}+a_{s+2}r^{s+2}+...\,,\nonumber\\
B(r)&=&b_t (r^t+b_{t+1}r^{t+1}+b_{t+2}r^{t+2}+...)\,.
\end{eqnarray}
There are three families of solutions as characterized by the powers of the first nonvanishing terms $(s,t)$ \cite{Stelle:1977ry}. We list properties and the free parameters of these families in Tab.~\ref{tab:ThreeFS}. We will not include the leading coefficient $b_t$ in parameter counting since it is determined by $B(\infty)=1$.  The remaining infinite set of coefficients are all determined, and we illustrate this up to some order in Appendix.~\ref{app:SeriesExp}.

\begin{table}[h]
\begin{center}
\caption{Properties and free parameters for three families of solutions.}
\vspace{1em}
\begin{tabular}{|c|c|c|c|c|}
\hline
&&&&
\\[-3mm]
$(s,t)$\,\, & \,\,behavior at $r=0$\,\, & \,\,generic CQG\,\, & \,\,$\beta=0$ CQG\,\, & GR
\\
&&&&
\\[-3.5mm]
\hline
$(0,0)$  & non-singular & $a_2, b_2$ & $b_2$ & none
\\
\hline
$(1,-1)$ & Schd-like &  $a_1, a_4, b_2$ & $a_1, a_4$ & \,\,$a_1$\,\,
\\
\hline
$(2,2)$ & \multirow{ 2}{*}{vanishing metric} & $a_2, a_5, b_3, b_4, b_5$ & $a_2, b_3, b_4$ 
&  \multirow{ 2}{*}{NA} 
\\
\cline{1-1}
\cline{3-4}
\,\,$(2,2)_E$\,\, & & $a_2, b_4$ & $a_2$  
& 
\\
\hline
\end{tabular}
\label{tab:ThreeFS}
\end{center}
\end{table}

The nonsingular $(0,0)$ family has only even power terms, and the increase in the number of free parameters in CQG indicates that Birkhoff's theorem no longer applies. The $(1,-1)$ family includes the Schd solution as a special case, as all vacuum solutions of GR automatically satisfy the field equations of CQG. It also includes other solutions with a horizon as found recently \cite{Lu:2015cqa}. 
Most interestingly CQG has a new type of solution, the $(2,2)$ family, that has no counterpart in GR. It is characterized by five free parameters \cite{Holdom:2002xy}, the same as the number of initial conditions needed to specify a solution of the field equations. At the origin all components of the metric $g_{\mu\nu}$ vanish. As we will see later there is a subclass of the $(2,2)$ family that deserves special attention. Like the $(0,0)$ family it has only even power terms in the series expansion and it also has the same number of free parameters. We denote this family by $(2,2)_E$ in the last row of Tab.~\ref{tab:ThreeFS}.   
In our exploration of asymptotically-flat thin-shell solutions in the fully nonlinear theory we shall focus on the $(0,0)$ and $(2,2)_E$ families. 

The presence of a smooth matter distribution does not affect the classification. But the expansion parameters of energy density and pressure do enter the determination of the remaining infinite set of coefficients in $A(r)$ and $B(r)$. For the $(0,0)$ family these quantities enter at ${\cal O}(r^4)$, while for the $(2,2)$ family they enter at ${\cal O}(r^{10})$ in generic CQG and ${\cal O}(r^8)$ in $\beta=0$ CQG.  

The leading behavior of four basic curvature invariants is also useful to characterize the different families of solutions. For comparison the Schd solution invariants in GR are\footnote{The $(1,-1)$ family in CQG has ($a_1=-1/2M$, $a_4=-a_1^4$ and $b_2=0$ gives Schd),
\begin{eqnarray}
&&R_{\mu\nu\rho\sigma}R^{\mu\nu\rho\sigma}=
C_{\mu\nu\rho\sigma}C^{\mu\nu\rho\sigma}=\frac{12}{a_1^2 r^6},\quad
R=\frac{3 }{2 a_1^2}\left(3 a_1^4+3 a_4-5 a_1 b_2\right),\nonumber\\
&&R_{\mu\nu}R^{\mu\nu}=\frac{9}{8a_1^4} \left(17 b_2^2a_1^2-6b_2a_1\left( a_1^4 +a_4\right)+9\left(a_1^4+a_4\right)^2\right).
\end{eqnarray}} 
\begin{eqnarray}
R=0,\,\,
R_{\mu\nu}R^{\mu\nu}=0,\,\,
R_{\mu\nu\rho\sigma}R^{\mu\nu\rho\sigma}=C_{\mu\nu\rho\sigma}C^{\mu\nu\rho\sigma}=48M^2/r^6\,.
\end{eqnarray}
The $(0,0)$ family has
\begin{eqnarray}
&&R=6(a_2-b_2),\,\,\,\,
R_{\mu\nu}R^{\mu\nu}=12(a_2^2-a_2b_2+b_2^2),\,\,\,\, \frac{}{}
R_{\mu\nu\rho\sigma}R^{\mu\nu\rho\sigma}=12(a_2^2+b_2^2),\nonumber\\
&&C_{\mu\nu\rho\sigma}C^{\mu\nu\rho\sigma}=\frac{(18 \beta (a_2^2-b_2^2)  + \Mp^2(a_2+2 b_2))^2}{300 \alpha ^2 }r^4\, ,
\end{eqnarray}
and the $(2,2)$ family has
\begin{eqnarray}\label{eq:curvature22E}
&&R=\frac{27 a_5 +a_2 \left(b_3(14 a_2 -2 b_3^2+10 b_4)-45 b_5\right)}{3 a_2^2 r},\quad
R_{\mu\nu}R^{\mu\nu}=\frac{12}{a_2^2r^8},\nonumber\\ 
&&R_{\mu\nu\rho\sigma}R^{\mu\nu\rho\sigma}=\frac{24}{a_2^2r^8},\quad
C_{\mu\nu\rho\sigma}C^{\mu\nu\rho\sigma}=\frac{(2 a_2 -2 b_4+b_3^2)^2}{3 a_2^2 r^4}\,.
\end{eqnarray}
Compared with the Schd singularity, the $(2,2)$ singularity is characterized by a stronger $1/r^8$ divergence for $R_{\mu\nu\rho\sigma}R^{\mu\nu\rho\sigma}$ but a weaker $1/r^4$ divergence for $C_{\mu\nu\rho\sigma}C^{\mu\nu\rho\sigma}$. These results for $R$ and $C^2$ reflect intricate cancellations that cause more singular terms to vanish. And in fact the Ricci scalar $R$ is not singular for the $(2,2)_E$ family. It approaches a negative constant $(-a_2^2+b_4^2-9a_2)/3a_2$ in generic CQG while it identically vanishes in $\beta=0$ CQG. Since the whole metric $g_{\mu\nu}$ vanishes with $r^2$, it is the inverse metric that drives the divergences in curvature invariants. With lowered indices the curvature tensor $R_{\mu\nu\alpha\beta}$ is regular and the Weyl tensor $C_{\mu\nu\alpha\beta}$ vanishes as $r^2$. 

In this paper we focus on solutions that are asymptotically-flat, which guarantees the weak field approximation at large distance. General solutions for the linearized field equations are as follows, with $A(r)=1+W(r)+{\cal O}(W^2)$ and $B(r)=1+V(r)+{\cal O}(V^2)$ \cite{Stelle:1977ry},
\begin{eqnarray}
\label{eq:linearS}
V(r)&=&\frac{2M}{r}+C_{0-}\frac{e^{-m_0r}}{r}+C_{0+}\frac{e^{m_0r}}{r}+C_{2-}\frac{e^{-m_2r}}{r}+C_{2+}\frac{e^{m_2r}}{r}\, ,\nonumber\\
W(r)&=&-\frac{2M}{r}+C_{0-}\frac{e^{-m_0r}}{r}(1+m_0r)+C_{0+}\frac{e^{m_0r}}{r}(1-m_0r)\nonumber\\
&&-\frac{1}{2}C_{2-}\frac{e^{-m_2r}}{r}(1+m_2r)-\frac{1}{2}C_{2+}\frac{e^{m_2r}}{r}(1-m_2r)\, .
\end{eqnarray}
$m_2^2=\Mp^2/2\alpha$ and $m_0^2=\Mp^2/6\beta$ are the masses of additional spin-0 and spin-2 degrees of freedom. Here we see that the linearized solution also has five free parameters, $M$, $C_{0\pm}$, $C_{2\pm}$. This is still true in the case $\alpha=3\beta$ of interest below where $m_2=m_0$, while for $\beta=0$ the spin-0 mode is decoupled leaving three parameters. The asymptotic flatness switches off the exponentially growing modes, i.e. $C_{0+}=C_{2+}=0$, and then $M$ is the physical mass. The solution at large $r$ approaches the Schd solution with two parameters $C_{0-}$, $C_{2-}$ characterizing exponentially small corrections, while for $\beta=0$ CQG there is one parameter $C_{2-}$. Clearly for $M\sim M_{\odot}$ and $m_0, m_2\sim\Mp$ and for radii where the weak field expansion is applicable, the exponentially small corrections are essentially invisible.  

This discussion leaves open the question of which of the families of solutions in CQG are actually realized as a response to matter distributions. This question for the linearized theory for different matter sources was studied analytically in \cite{Lu:2015psa}. It was found that the exponentially small terms are always nonzero and encode information about the source. As for the fully nonlinear solutions, an example of a vacuum $(2,2)$ solution was obtained numerically in \cite{Holdom:2002xy}, as were high curvature $(0,0)$ solutions for incompressible matter. This leaves much room for a more systematic study of asymptotically-flat solutions that are sourced by matter in the fully nonlinear theory. In fact our results will significantly differ from the corresponding attempt in \cite{Lu:2015psa} along these lines.

The simplest delta-function source in the fully nonlinear theory is a spherical thin-shell of radius $\sl$ \cite{Geroch:1987qn}. In Sec.~\ref{sec:thin-shell} we describe thin-shell models in CQG and our numerical strategy to search for asymptotically-flat solutions in the multi-dimensional parameter space. We present our numerical results in Sec.~\ref{sec:numsolution}, where the $(2,2)_E$ and $(0,0)$ families together provide a comprehensive picture of the horizonless thin-shell solutions in the fully nonlinear theory. In Sec.~\ref{sec:scaling} we find an interesting $M$ scaling behavior of the interior region of the $(2,2)_E$ solutions for large $M$.

\subsection{Thin-shell models}
\label{sec:thin-shell}

For generic CQG we follow the setup of a thin-shell source in \cite{Lu:2015psa}, which we refer to as the TS1 model. It is described by the stress tensor $T_{\mu\nu}=\mathrm{diag}(T_{tt},\,T_{rr},\,T_{\theta\theta},\,T_{\theta\theta}\sin^2\theta)$ with
\begin{eqnarray}
\label{eq:thin-shell1}
T_{tt}(r)=B(r)\rho(\sl)\delta(r-\sl),\quad
T_{rr}(r)=0\,.
\end{eqnarray}
$T_{\theta\theta}(r)$ is then determined by the only nontrivial conservation law $\nabla^\mu T_{\mu r}=0$,
\begin{eqnarray}\label{eq:MomC1}
T_{\theta\theta}(r)=\frac{r^3B'(r)}{4B^2(r)}T_{tt}(r)\,.
\end{eqnarray}
We can also set $T_{\theta\theta}(r)=r^2\tilde{p}(\sl)\delta(r-\sl)$. The vacuum solutions inside and outside the shell are matched at $r=\sl$ with five conditions, where $A$, $A'$, $B'/B$, $B''/B$ are continuous while $A''$ has a jump as determined by the energy density on the shell,
\begin{eqnarray}
\label{eq:jump1}
A''_{\mathrm{out}}(\sl)-A''_{\mathrm{in}}(\sl)=2\pi A^3(\sl) \sl \rho(\sl)\frac{(\alpha -3 \beta )B' (\sl)\sl-2 (\alpha +6 \beta )B(\sl) }{9 \alpha  \beta  B(\sl)}\,.
\end{eqnarray}
The derivatives are respect to $r$, evaluated at $\sl$. The value of $B$ itself is trivially continuous at the shell.

If the shell moves from one radius to another as a function of time, the stress tensor needs to satisfy the energy conservation law, $\nabla^\mu T_{\mu t}=0$, in addition to (\ref{eq:MomC1}). The time dependence can be factorized out as $d\sl/dt$ and the conservation law is reduced to a constraint on how the shell energy density depends on the shell radius, 
\begin{eqnarray}\label{eq:EngC1}
\frac{d }{d \sl}\sigma(\sl)+\frac{2}{\sl}\left(\sigma(\sl)+p(\sl)\right)=0
.\end{eqnarray}
Here $\sigma(\sl)\equiv \sqrt{A(\sl)}\rho(\sl)$ and $p(\sl)\equiv \sqrt{A(\sl)}\tilde{p}(\sl)$. The factor of $\sqrt{A(\sl)}$ is related to defining a proper unit length along the radial direction and it brings agreement with the Israel junction condition \cite{Israel:1966rt}.  The conservation law (\ref{eq:MomC1}) becomes\footnote{This relation does not apply to the thin-shell in GR because a discontinuous $A(r)$ makes the form of the conservation law different.}
\begin{align}
\label{cons1}
{p(\sl)\over\sigma(\sl)}={\sl B'(\sl)\over4B(\sl)}.
\end{align}
Combining (\ref{eq:EngC1}) and (\ref{cons1}) gives
\begin{eqnarray}\label{eq:EngC2}
\frac{d }{d \sl}\sigma(\sl)+\frac{2}{\sl}\sigma(\sl)\left(1+\frac{\sl B'(\sl)}{4B(\sl)}\right)=0\,.
\end{eqnarray}
When we study numerical solutions in Sec.~\ref{sec:numsolution}, we shall indeed confirm that solutions at different $\sl$ with the same physical mass $M$ satisfy this relation.  

We find asymptotically-flat thin-shell solutions by the shooting and matching method: 
\begin{enumerate}[1)]
\item shooting from the outside with small deviations from the Schd solution with a fixed $M$; 
\item shooting from the inside using the known series expansion to determine initial conditions at a $r_0$ close to 0;
\item adjusting parameters to match at $r=\sl$ the four continuous quantities $A$, $A'$, $B'/B$, $B''/B$;
\item identify the energy density $\rho(\sl)$ on the shell from the jump of $A''$ by (\ref{eq:jump1}).  
\end{enumerate}

In principle linearized solutions (\ref{eq:linearS}) could be used as a starting point for shooting from the outside. But in practice a rather large $r$ is required to ensure the validity of the linear approximation and numerical errors are easily accumulated. Fortunately as the exact solution in GR, the Schd solution provides a better approximation in the small curvature region, and we can consider small deviations from that in the initial conditions. Then we immediately have a precise definition of $M$ for a numerical solution. Thus $M$ and $\sl$ are taken as inputs to specify a solution with $\rho(\sl)$ determined by the solution through (\ref{eq:jump1}).

There are four quantities $A$, $A'$, $B'/B$, $B''/B$ that need matching at $r=\sl$. Shooting from the outside can still be considered to have two free parameters as suggested by the linearized solutions in (\ref{eq:linearS}). Shooting from the inside has the parameters of the series expansion and from Tab.~\ref{tab:ThreeFS}, both $(0,0)$ and $(2,2)_E$ families have two parameters. Thus in either case there are the needed four parameters for the matching at $r=\sl$. In \cite{Lu:2015psa} this type of parameter counting was used to suggest that a $(0,0)$ solution exists even when the matter shell is well within the would-be horizon. However as we will see in Sec.~\ref{sec:numsolution}, such an argument is far from sufficient to ensure that any of these solutions exist in the nonlinear regime.

In addition to the numerical study of generic CQG we shall also develop $\beta=0$ CQG, as it leads to a simpler numerical problem. Unfortunately it is less straightforward to set up a thin-shell model in this case. When $\beta=0$ the original $tt$ and $rr$ field equations need to be rewritten in an equivalent form to make the differential order manifest, namely, two second order differential equations \cite{Lu:2015psa}. In analogy with the thin-shell TS1 model we want $A$, $B'/B$ to be continuous while $A'$ jumps across the shell. This leads to the TS2 model where the stress tensor takes the form
\begin{eqnarray}
\label{eq:thin-shell2}
T_{rr}(r)=A(r)p_r(\sl)\delta(r-\sl),\,\,\,
T_{tt}(r)=\frac{B(r)}{A(r)}\Big(X(r)T_{rr}(r)+Y(r)T'_{rr}(r)\Big),
\end{eqnarray}
with $X=(6B+r B'-2rB \,A'/A)/(2B-rB')$, $Y=2rB/(2B-rB')$. $T_{\theta\theta}(r)$ is again determined by the conservation law $\nabla^\mu T_{\mu r}=0$ which can be reduced to
\begin{eqnarray}
\label{eq:Tthth2}
T_{tt}(r)/B(r)=2T_{\theta\theta}(r)/r^2+T_{rr}(r)/A(r)\,.
\end{eqnarray}
The existence of a radial pressure and the fact that both $T_{tt}$ and $T_{\theta\theta}$ include a derivative of the delta-function makes the physical interpretation of the TS2 model somewhat more difficult than the TS1 model.  

Asymptotically-flat solutions in $\beta=0$ CQG can now be found in a similar way. For a given $(M,\sl)$ the metric functions are obtained by shooting from the outside with one parameter, shooting from the inside with the $(0,0)$ or $(2,2)_E$ families with one parameter, and then matching the values of ($A, B'/B$) at $r=\sl$. The jump for $A'$ then determines the shell property $p_r(\sl)$ by
\begin{eqnarray}
\label{eq:jump2}
A'_{\textrm{out}}(\sl)-A'_{\textrm{in}}(\sl)=-8\pi p_r(\sl)\frac{A^3(\sl) \sl^2 B(\sl) }{\alpha(2B(\sl)-\sl B'(\sl))}\,.
\end{eqnarray}

Finally we can make a comparison to GR. The thin-shell model in GR is similar to TS1, $T_{tt}(r)\sqrt{A(r)}/B(r)=\sigma(\sl)\delta (r-\sl)$, $T_{rr}(r)=0$, $T_{\theta\theta}(r)\sqrt{A(r)}/r^2=p(\sl)\delta (r-\sl)$. Here $p(\sl)/\sigma(\sl)=\pi\sigma(\sl)\sl/(\Mp^2-4\pi\sigma(\sl)\sl)$ and it is a jump in $A$ that is related to the shell,
\begin{eqnarray}
\label{eq:jump3}
A^{-1/2}_{\mathrm{out}}(\sl)-A^{-1/2}_{\mathrm{in}}(\sl)=-\frac{4\pi\sigma(\sl)\sl}{\Mp^2}
\,.\end{eqnarray}
This is all that needs to be determined according to Birkhoff's theorem. The matching of other continuous quantities in CQG, which enables a rich structure in the solution space, is simply absent in GR. 
 
\subsection{Asymptotically-flat thin-shell solutions in $(0,0)$ and $(2,2)_E$ families}
\label{sec:numsolution}

In this section we present numerical solutions in the fully nonlinear theory with the thin-shell model TS1 (\ref{eq:thin-shell1}) in generic CQG and with the TS2 (\ref{eq:thin-shell2}) model in $\beta=0$ CQG. Hereafter we set $\Mp=1$. For generic CQG we shall pick in particular $m_2=m_0=1$ ($\alpha=3\beta=1/2$), and for $\beta=0$ CQG we take $m_2=1$. With the shooting and matching method described in Sec.~\ref{sec:thin-shell}, we search for asymptotically-flat solutions in both the $(0,0)$ and $(2,2)_E$ families for selected pairs of $(M, \sl)$. In generic CQG ($\beta=0$ CQG) we do parameter scans in each of the 2D (1D) spaces that characterizes both shooting from the inside and the outside, and then determine the solution by the 4 (2) matching conditions at $r=\sl$. For $\beta=0$ CQG we shall be able to find solutions at significantly larger values of $M$.

\begin{figure}[!h]
  \centering%
{ \includegraphics[height=5.05cm]{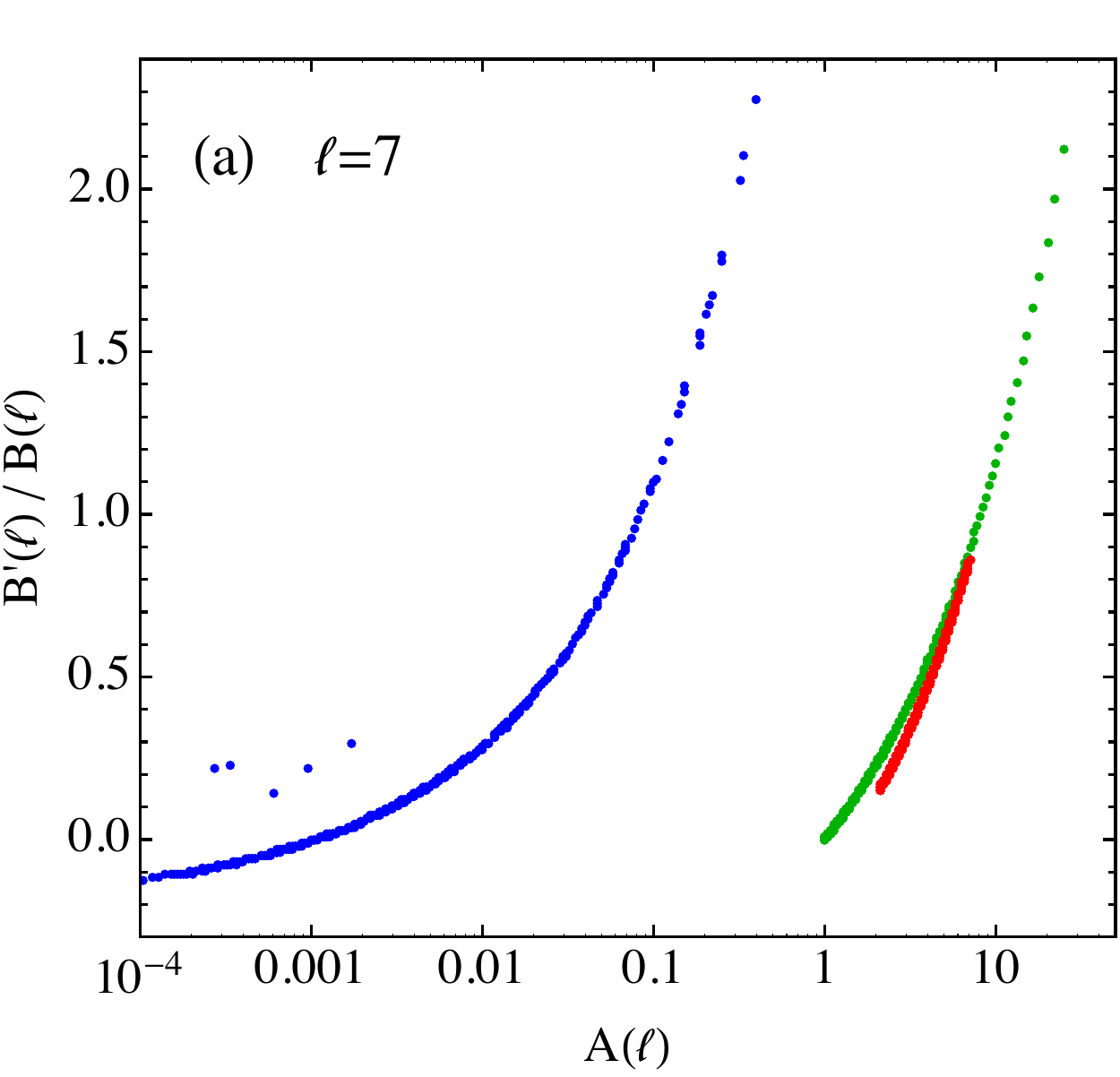}}\,\,
{ \includegraphics[height=5cm]{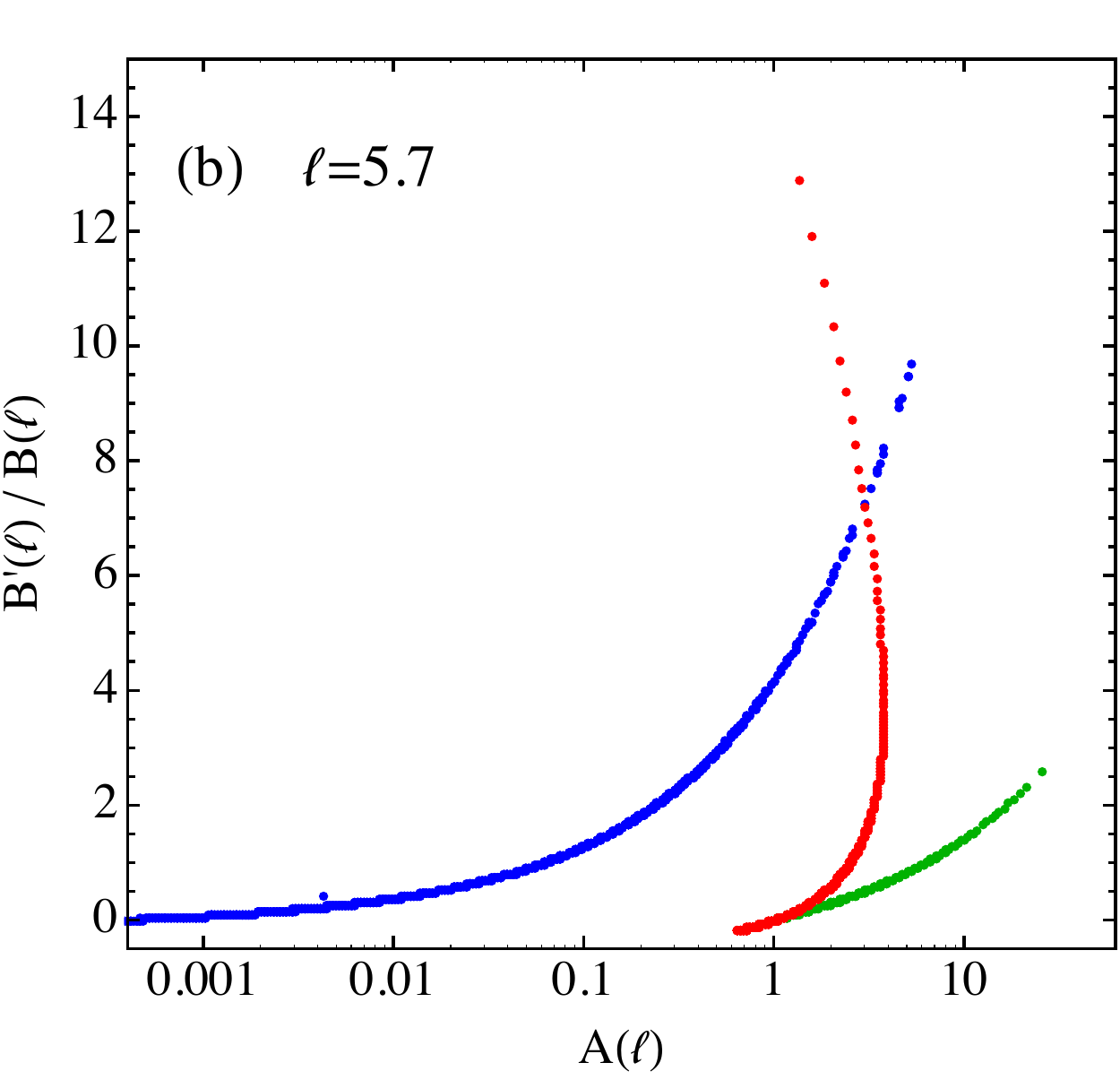}}\,\,
{ \includegraphics[height=5cm]{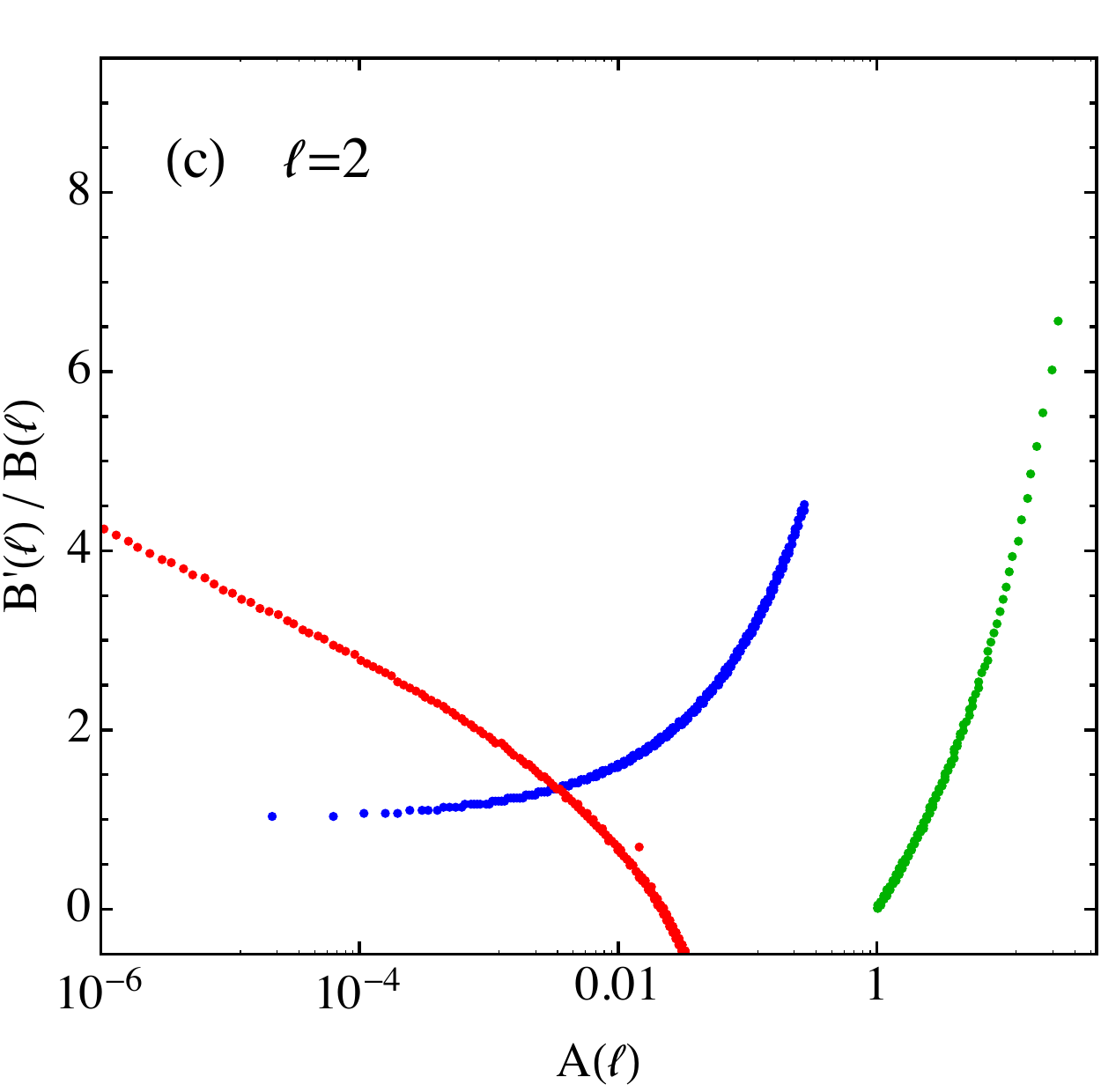}}
\caption{\label{fig:MatchM3} The matching on the $A(\sl)$ vs $B'(\sl)/B(\sl)$ plane at $\sl=2, 5.7, 7$ for $M=3$ in $\beta=0$ CQG. The red dots represent a one parameter scan from the outside shooting, while green and blue dots are one parameter scans from the inside shooting for the $(0,0)$ and $(2,2)_E$ families respectively.
}
\end{figure}

For illustration we start with small $M$ in $\beta=0$ CQG. Fig.~\ref{fig:MatchM3} shows how the matching works at different $\sl$  for $M=3$ on the $A(\sl)$ vs $B'(\sl)/B(\sl)$ plane. For there to be a solution the red dots must intersect with either the green or blue dots, corresponding to the $(0,0)$ and $(2,2)_E$ families respectively. When the shell radius $\sl$ is far outside the would-be horizon $r_H$ there are only solutions in the $(0,0)$ family as shown in Fig.~\ref{fig:MatchM3}(a). When $\sl$ approaches $r_H$, solutions in the $(2,2)_E$ family start to appear, and within a range we may have solutions in both families as in Fig.~\ref{fig:MatchM3}(b). Pushing the shell further inside $r_H$, $(0,0)$ solutions no longer exist and $(2,2)_E$ solutions take over as shown in Fig.~\ref{fig:MatchM3}(c). From this it is clear that a simple counting of parameters is far from sufficient to establish that a particular solution exists. The numerical analysis is essential to determine which of the possible interior behaviors is the correct one for a given $\sl/r_H$. When $M$ is larger a small gap opens up around $\sl\sim r_H$ where neither type of solution can be found, although whether this is just due to a numerical limitation is not clear.

Finding solutions in generic CQG corresponds to finding the intersection of two 2D surfaces in a four dimensional parameter space. That such an intersection even exists is nontrivial. We find that when $\sl$ is not much smaller than $r_H$, the intersection successfully determines the four parameters used in the matching. This is true for both $(0,0)$ and $(2,2)_E$ solutions. When $\sl$ becomes small compared to $r_H$ then a new phenomenon occurs. The 2D surface from the outside shooting becomes closer and closer to just a 1D line. The intersection with the other 2D surface still occurs, but now because of  the finite numerical accuracy and the nearly 1D line, the parameters governing the shooting from the outside are no longer determined. We shall make use of the conservation law (\ref{eq:EngC2}) to help pin down the correct solution at small $\sl$.

\begin{figure}[!h]
  \centering%
{ \includegraphics[width=7.5cm]{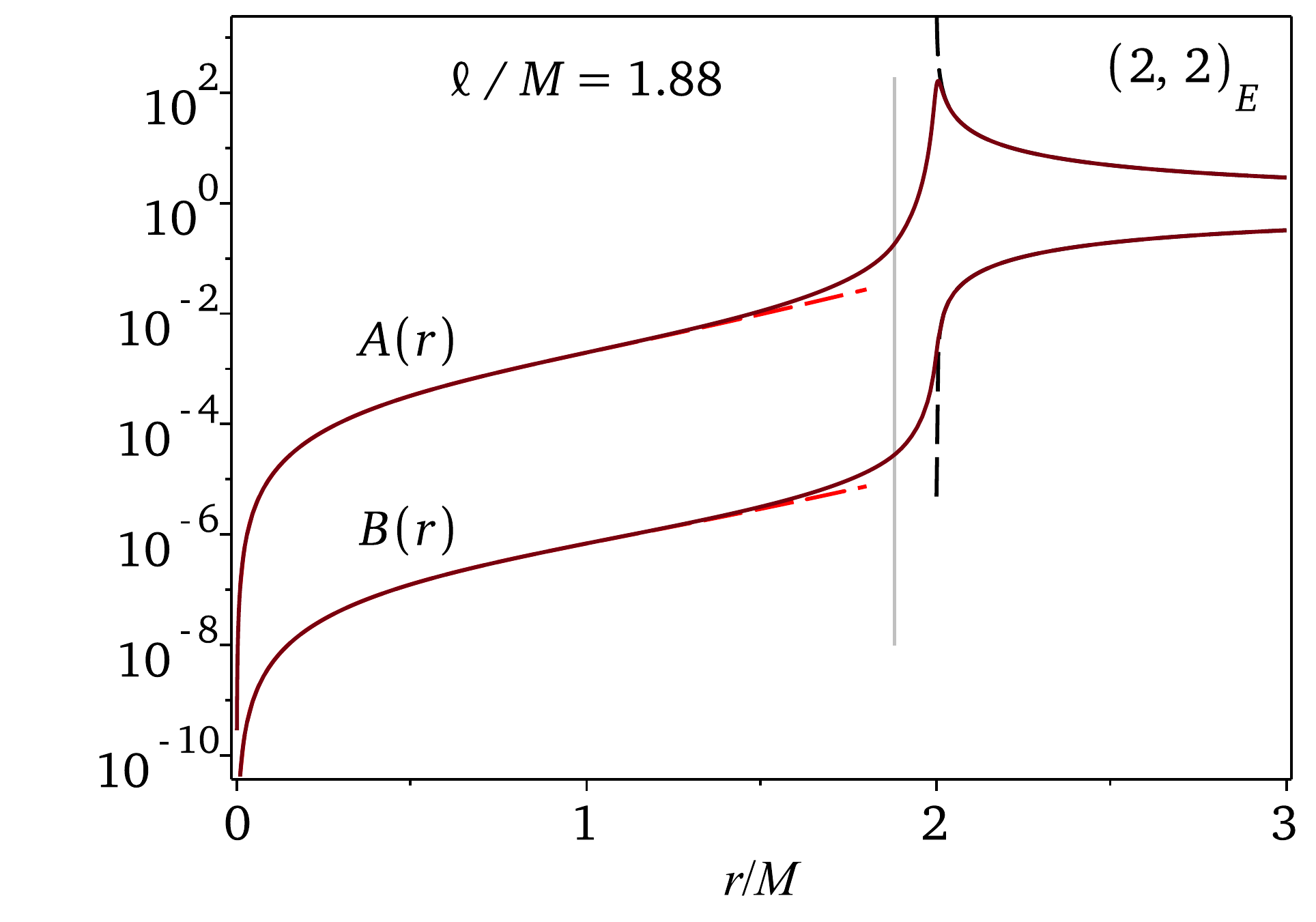}}\,\,
{ \includegraphics[width=7.5cm]{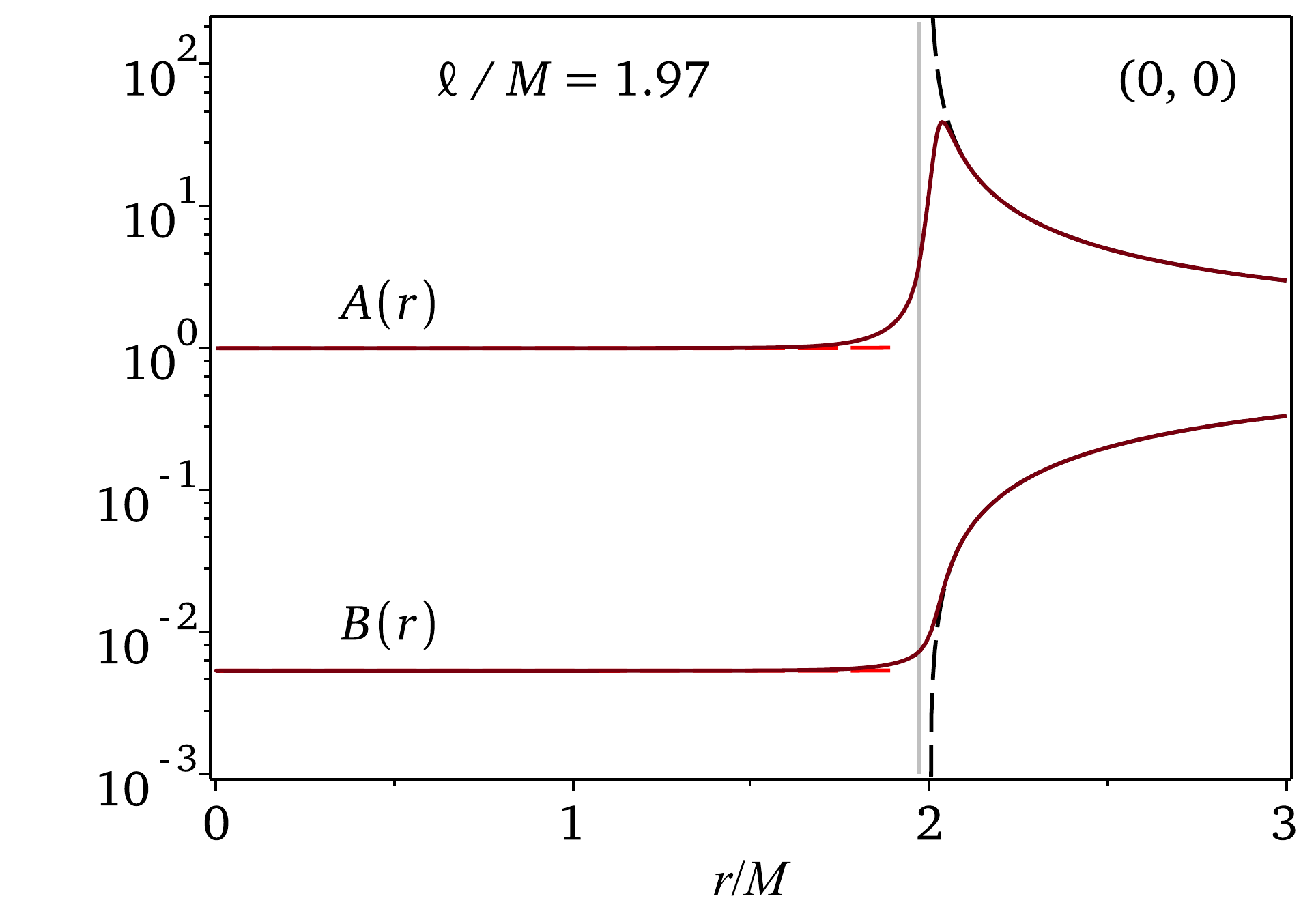}}
\caption{\label{fig:ABSM10}  Numerical solutions for $A(r)$ and $B(r)$ for $M=10$ in generic CQG. Left, the $(2,2)_E$ solution with shell radius $\sl/M=1.88$. Right, the $(0,0)$ solution with $\sl/M=1.97$. The vertical gray lines denote these $\sl$'s. The black dashed lines denote the Schd solution while the colored dashed lines denote the series expansion solutions.}
\end{figure}

Fig.~\ref{fig:ABSM10} shows examples of $A(r)$ and $B(r)$ solutions for generic CQG for $M=10$. It shows a $(2,2)_E$ solution with a shell radius $\sl=18.8$ and a $(0,0)$ solution with a shell radius $\sl=19.7$. Between these two $\sl$'s we are unable to find solutions of either type. Note that the $(0,0)$ solution has $\sl<r_H$ where a black hole would have already formed in GR. (This was also found for a different matter source in \cite{Holdom:2002xy} and it is also seen in Fig.~\ref{fig:MatchM3}(b)). Both solutions closely match the Schd solution for $r>r_H$, for a range of $r$ that is finite.\footnote{The range is finite because one cannot ensure that the exponentially growing modes are identically zero in a numerical approach.} The series expansions also agree well with the numerical solutions for most of the interior.

\begin{figure}[!h]
  \centering%
{ \includegraphics[width=7.5cm]{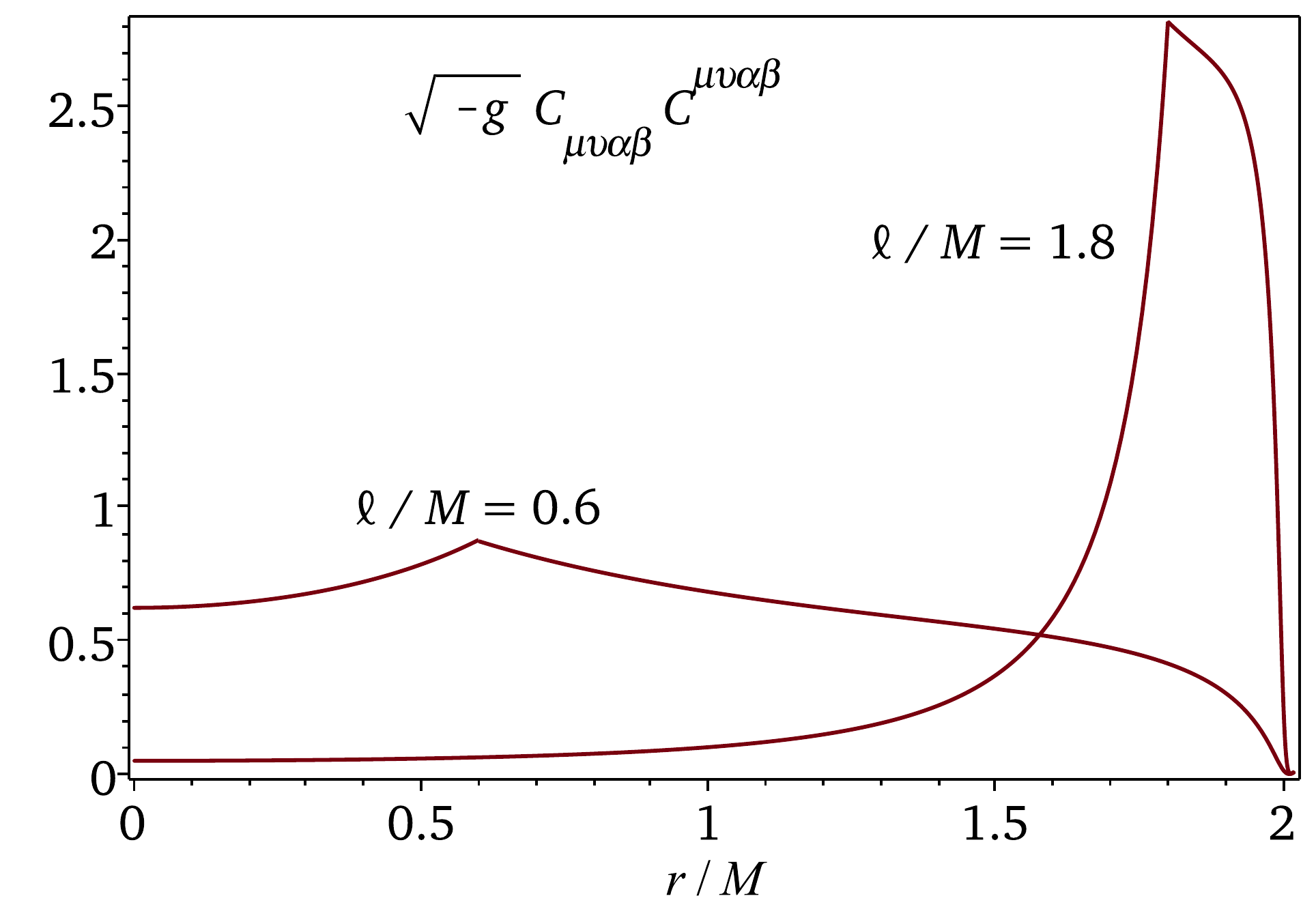}}\quad
{ \includegraphics[width=7.5cm]{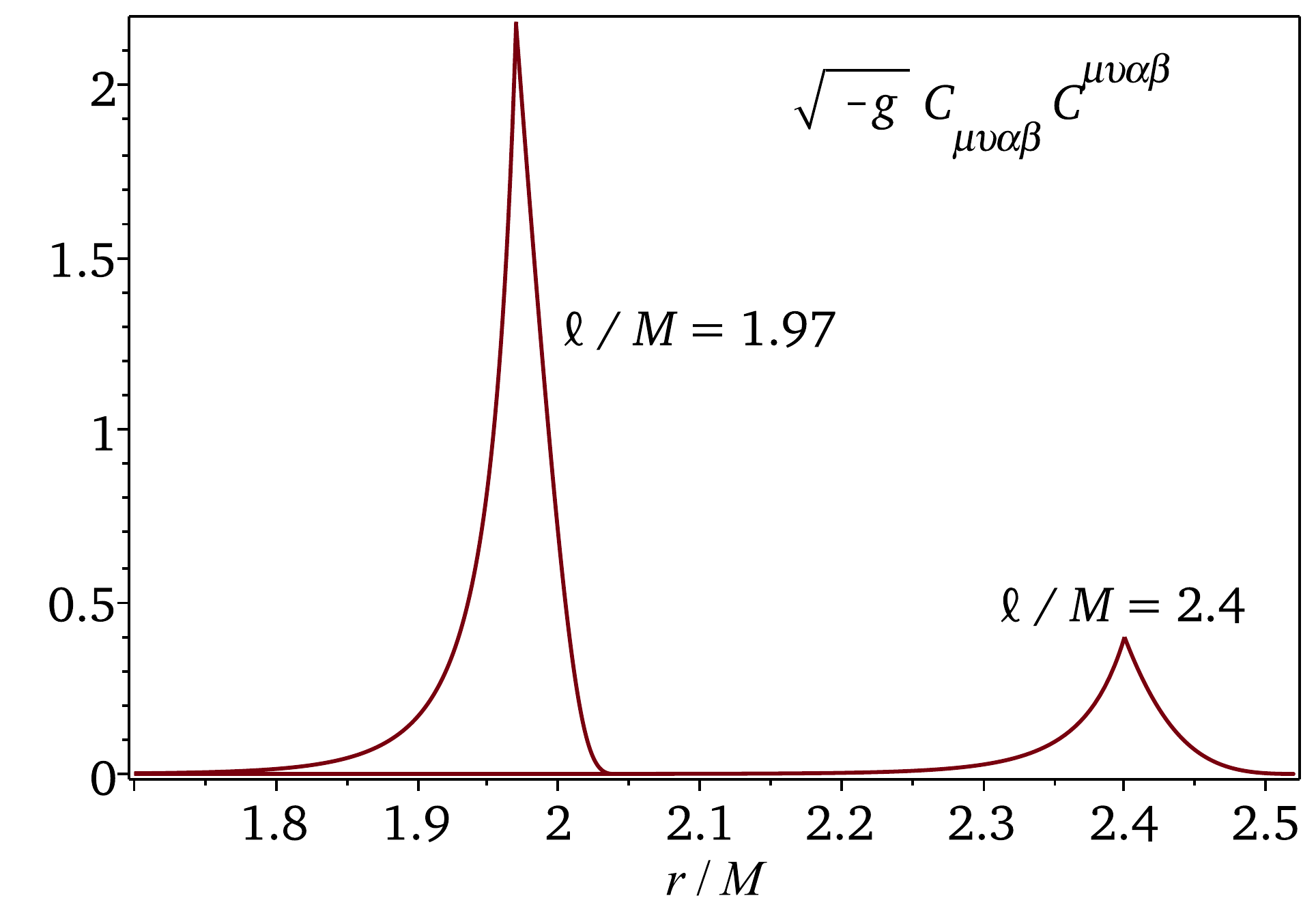}}\quad\quad
\caption{\label{fig:sqrtgC2}  The Weyl term $\sqrt{-g}C_{\mu\nu\rho\sigma}C^{\mu\nu\rho\sigma}$ in the CQG Lagrangian for the $(2,2)_E$ (left) and $(0,0)$ (right) solutions for $M=10$ in generic CQG.
}
\end{figure}

In the interior of the $(2,2)_E$ solution, the sharp decline of $A(r)$ and $B(r)$ implies a shrinking 4-volume. Also the curvatures become super-Planckian and keep growing towards a timelike singularity at the origin as in (\ref{eq:curvature22E}). But since $\sqrt{-g}\sim r^4$, the Lagrangian density for these solutions is finite in the whole spacetime. Fig.~\ref{fig:sqrtgC2} shows the $r$ dependence of the term $\sqrt{-g}C_{\mu\nu\rho\sigma}C^{\mu\nu\rho\sigma}$ for the $(2,2)_E$ and $(0,0)$ solutions. In the $(2,2)_E$ case we see inner structure of the 2-2-hole that is dependent on the shell location. In the $(0,0)$ case curvature invariants peak at the location of the shell. Away from the shell they approach the exterior Schd behavior or the interior flat spacetime exponentially quickly.

\begin{figure}[!h]
  \centering%
{ \includegraphics[width=7.5cm]{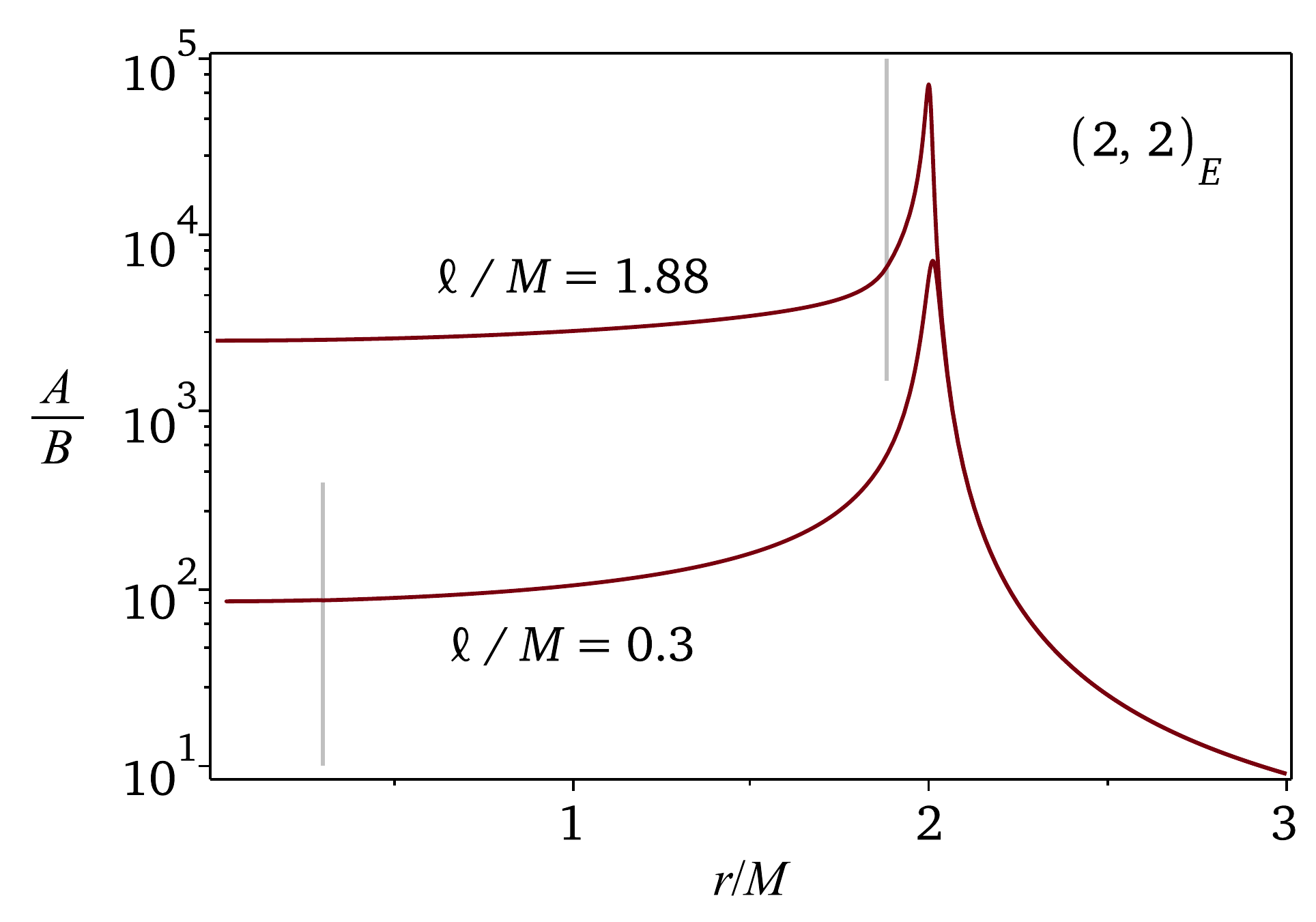}}
{ \includegraphics[width=7.5cm]{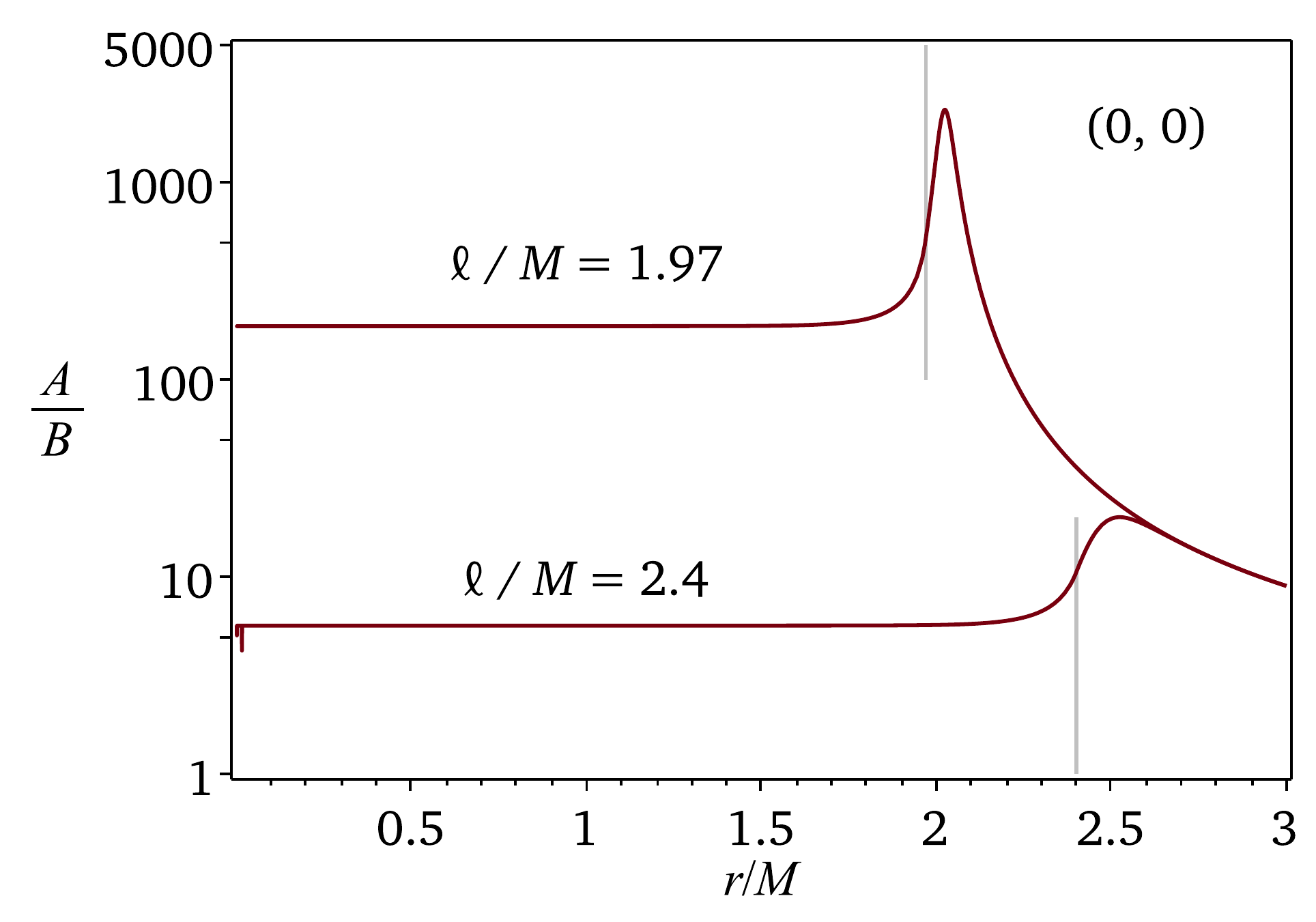}}\quad\quad\\
\quad\,\,{ \includegraphics[width=7.4cm]{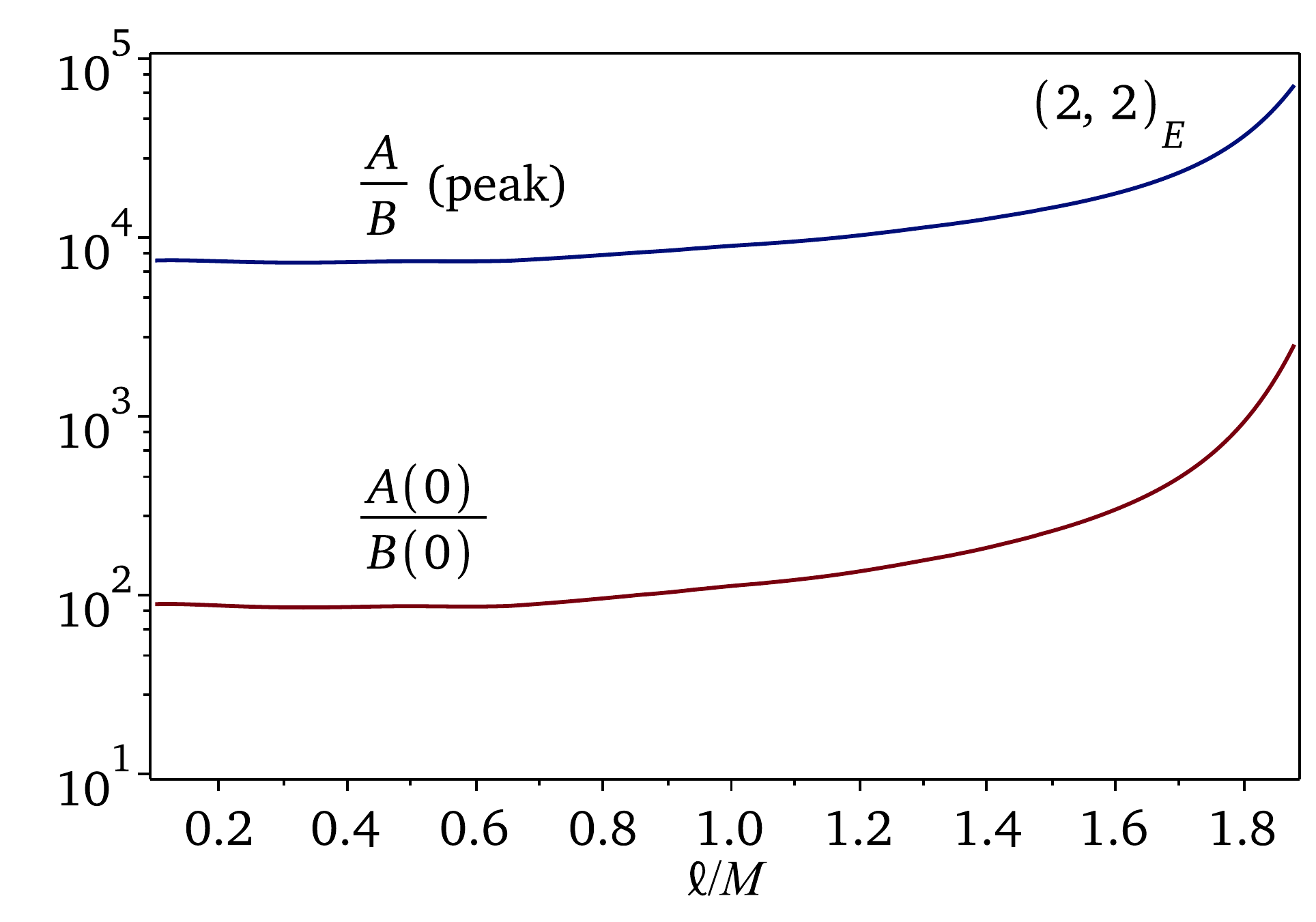}}\,\,
{ \includegraphics[width=7.4cm]{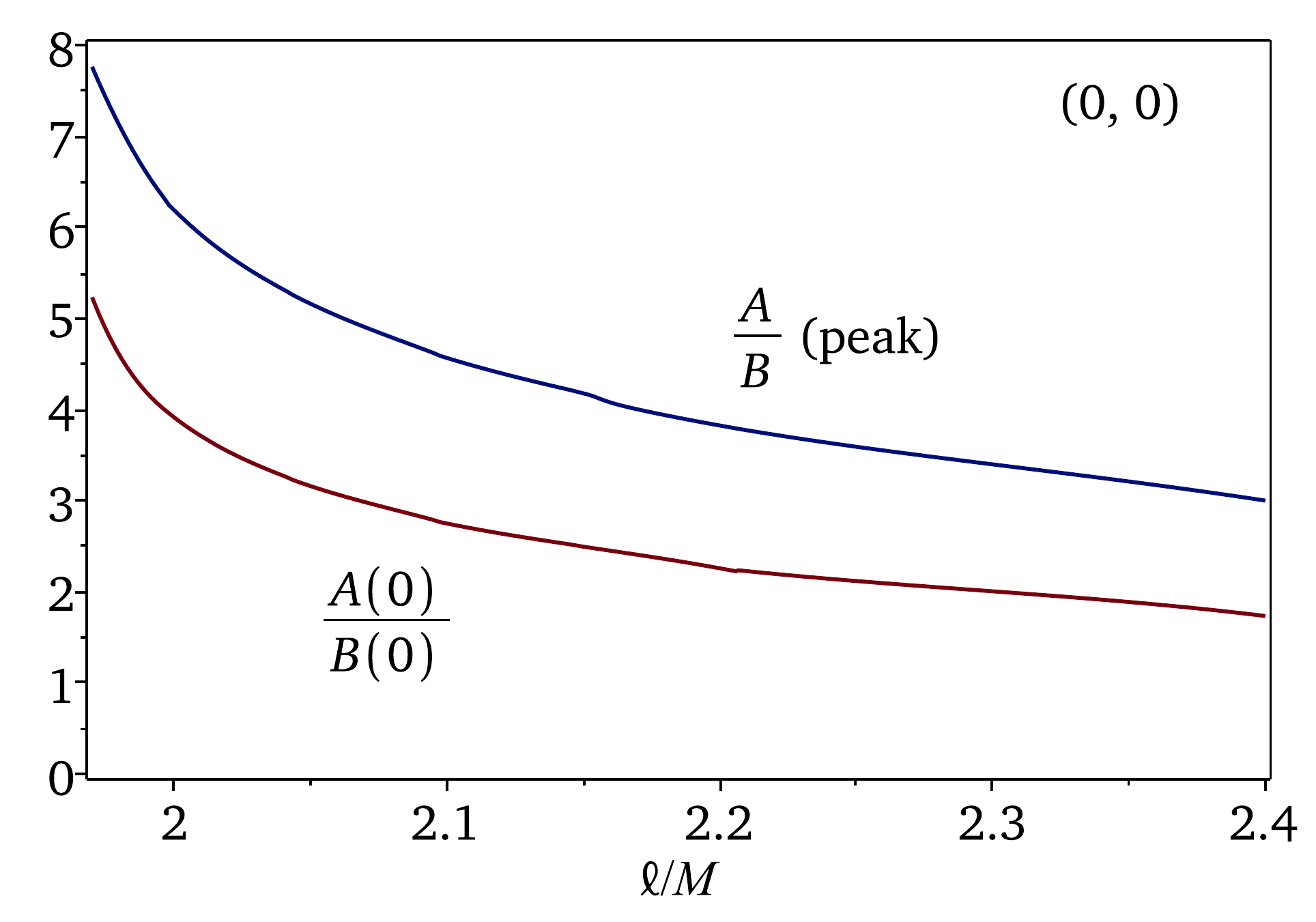}}\, \,
\caption{\label{fig:ABRM10} 
Upper, the ratio $A(r)/B(r)$ with $M=10$ at different $\sl$ for the $(2,2)_E$ and $(0,0)$ solutions in generic CQG. Lower, the peak value and the interior constant value of the ratio as function of $\sl$. }
\end{figure}

Another quantity of interest is $A(r)/B(r)$ which defines the tortoise coordinate, $d r_*/dr=\sqrt{A/B}$. The integration of the tortoise coordinate determines the coordinate time $\Delta t$ for light to traverse a certain radial distance $\Delta r$.  Fig.~\ref{fig:ABRM10}(upper) shows $A(r)/B(r)$ for the $(2,2)_E$ and $(0,0)$ solutions at various $\sl$. For either case the ratio reaches a peak around the radius where the deviation from the Schd solution occurs. Inside the peak it decreases and approaches the $r=0$ value, which is $a_2/b_2$ and $1/b_0$ for the $(2,2)_E$ and $(0,0)$ solutions respectively. Fig.~\ref{fig:ABRM10}(lower) shows the peak value and the $r=0$ value of $A(r)/B(r)$ as functions of the shell radius. The corresponding plots for $\beta=0$ CQG are roughly similar.

\begin{figure}[!h]
  \centering%
{ \includegraphics[width=7.5cm]{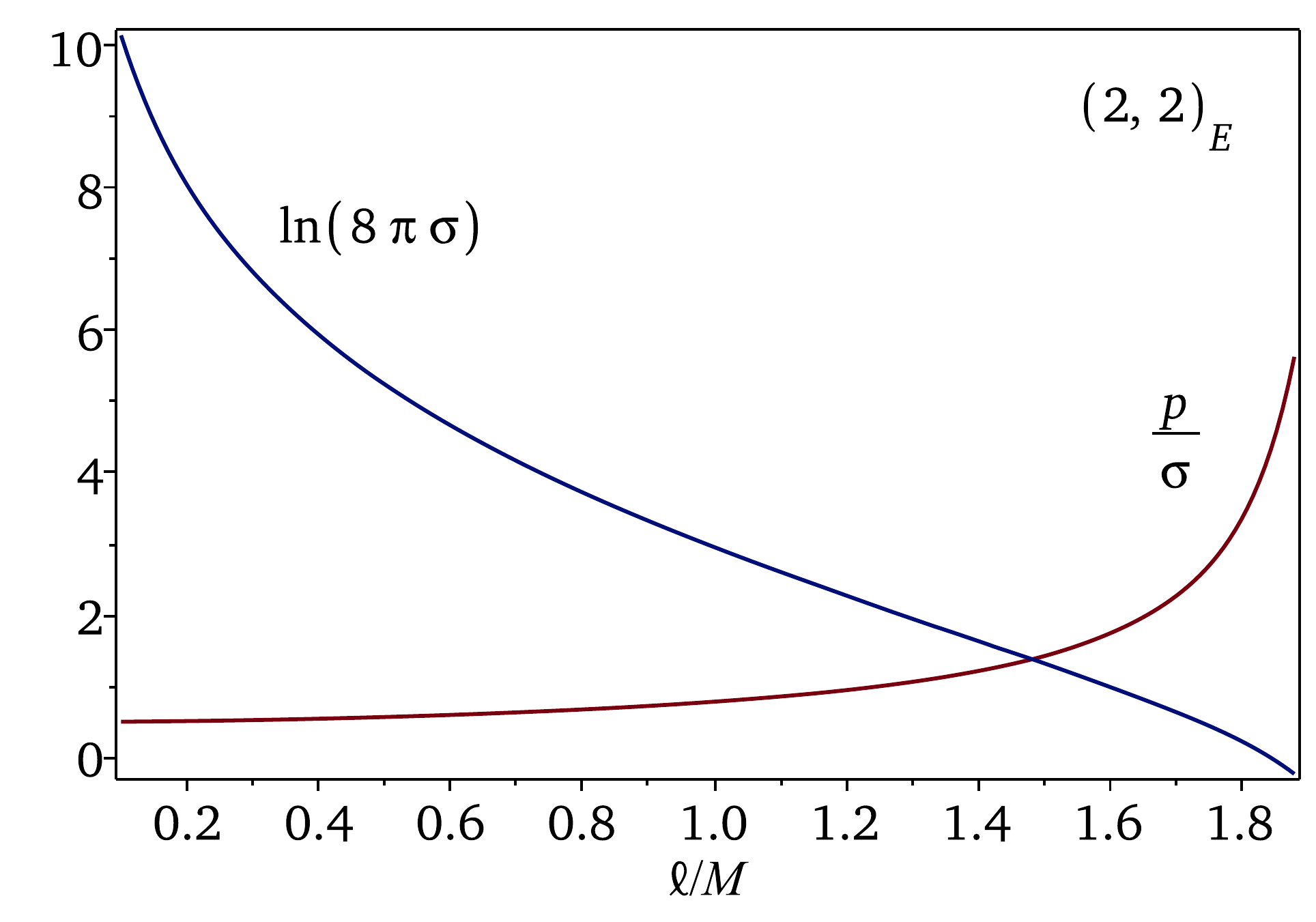}}\,\,\,\,
{ \includegraphics[width=7.5cm]{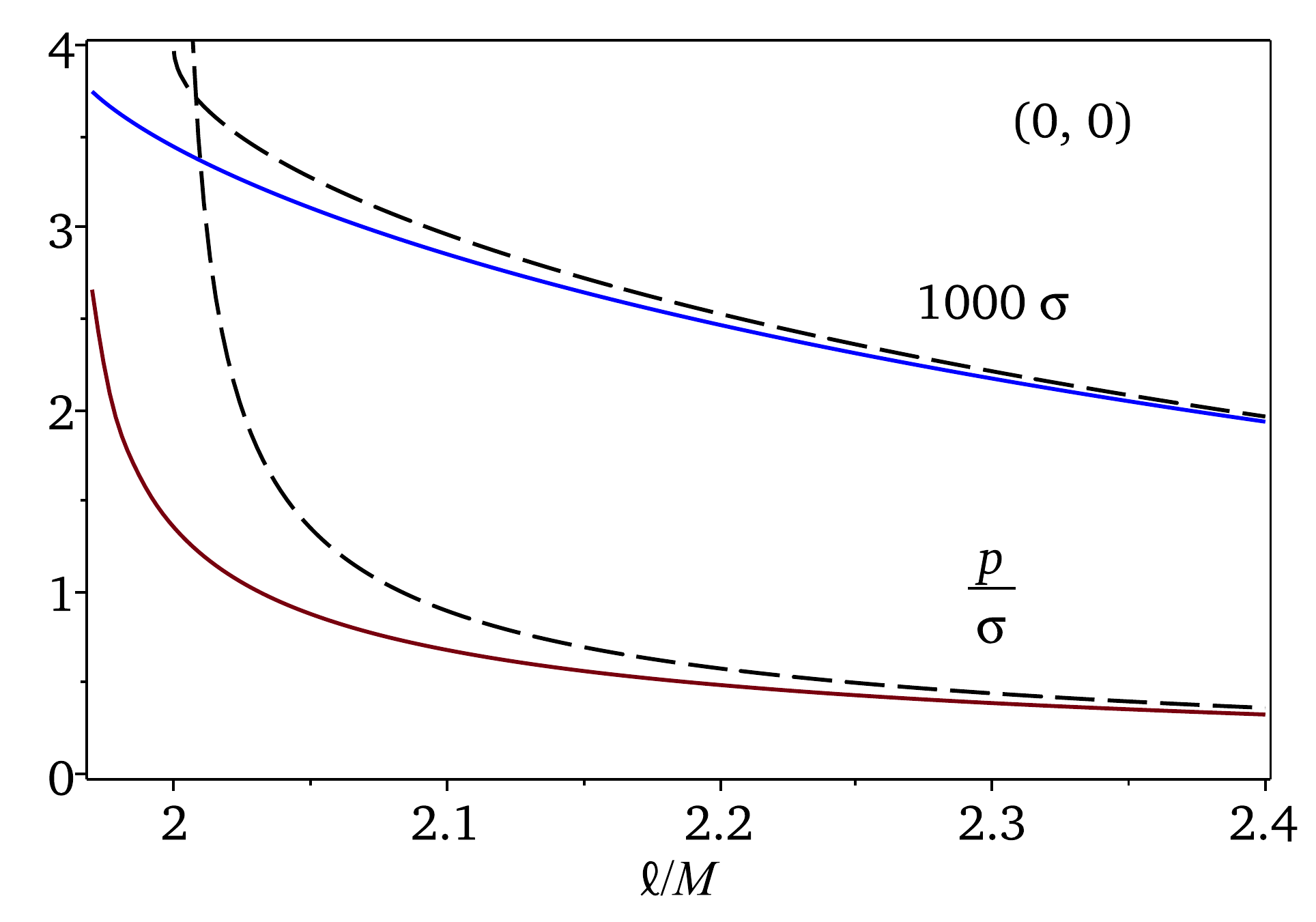}}
\caption{\label{fig:shellPM10} 
The energy density $\sigma(\sl)$ and the ratio $p(\sl)/\sigma(\sl)$ with $M=10$ for the $(2,2)_E$ and $(0,0)$ solutions in generic CQG. }
\end{figure}

With numerical solutions for $A(r)$ and $B(r)$ in generic CQG we can obtain the conserved energy density $\sigma(\sl)$ and the pressure $p(\sl)$ from the relations (\ref{eq:jump1}) and (\ref{cons1}). These quantities are displayed in Fig.~\ref{fig:shellPM10} and we have confirmed that they are consistent with the conservation law (\ref{eq:EngC2}).\footnote{As we have mentioned before,  when $\sl$ is small compared to $r_H$ we actually use (\ref{eq:EngC2}) to help determine the solution.} This provides a nontrivial check of our numerical results. For the $(2,2)_E$ solution we see that for $\sl\gtrsim1.25M$ the dominant energy condition $|p(\sl)|\leq\sigma(\sl)$ is violated. At small $\sl$, $\;\sl B'(\sl)/4B(\sl)\approx 1/2$ and so $\sigma(\sl)\sim 1/\sl^3$. This corresponds to $T_{tt}$ and $T_{\theta\theta}$ scaling like $1/\sl^2$. For the $(0,0)$ solution, $p(\sl)/\sigma(\sl)$ quickly drops below unity for increasing $\sl$, while $\sigma(\sl)$ approaches the prediction of the thin-shell model in GR (\ref{eq:jump3}). This in turn approaches the weak gravity limit $\sigma(\sl)= M/4\pi\sl^2$ at large $\sl$.

In summary the thin-shell model and the special $(2,2)_E$ family nicely illustrate the complementarity between the novel $(2,2)$ solutions and the star-like $(0,0)$ solutions in describing the high and low compactness respectively. Also apparent are some similarities that $(2,2)_E$ solutions have with $(0,0)$ solutions, similarities that are not shared with black holes.

\subsection{Large $M$ and scaling behavior}
\label{sec:scaling}

To make a connection with astrophysical black hole candidates we need to know the general behavior of 2-2-holes for enormously larger masses, $M\sim M_{\odot}\sim 10^{38}\Mp$. To get some flavor we obtained $(2,2)_E$ solutions for $M\in[10,10^4]$ in $\beta=0$ CQG and also for $M=15$ and 20 in generic CQG, for some values of $\sl$. These larger $M$ solutions continue to have features similar to what we have presented above. But the peak in $A(r)/B(r)$ continues to grow and the deviation from the Schd solution occurs closer and closer to $r_H$. In Sec.~\ref{sec:QNMs} we shall study this nontrivial behavior in the peak region more closely.

The study of these larger $M$ solutions was sufficient to uncover an interesting result for the interior of a 2-2-hole. This interior, where $r$ is at least somewhat smaller than $r_H$, is governed by a simple scaling law. For a given $\sl/M$ and for different masses $M$ and $\varrho M$, the metric functions $A(r)$, $B(r)$ and any curvature invariant $I(r)$ are related as follows,    
\begin{eqnarray}\label{eq:22scaling}
A_M(r)=\varrho^2 A_{\varrho M}(r\varrho),\,\,
B_M(r)=\varrho^2 B_{\varrho M}(r\varrho),\,\,
I_M(r)=I_{\varrho M}(r\varrho)\,.
\end{eqnarray}
With increasing $M$ the scaling region expands so that it applies to $r/M$ closer and closer to $r_H/M=2$. This scaling is in contrast to the behavior of the Schd solution which has $A_M(r)= A_{\varrho M}(r\varrho)$, $B_M(r)= B_{\varrho M}(r\varrho)$ and $I_M(r)=\varrho^{2n}I_{\varrho M}(r\varrho)$, where the dimension of $I(r)$ is $2n$.

So given the $\sl$ dependent 2-2-hole solutions at one $M$, we now know the 2-2-hole interior solutions for any large $M$. We find that $\rho$ and $p_r$ that appear in the two thin-shell models scale as $M$, as can be seen by applying the scaling of (\ref{eq:22scaling}) to the jump conditions (\ref{eq:jump1}) and (\ref{eq:jump2}) respectively. The conserved energy density $\sigma$ then scales as $M^0$ and is thus only a function of $\sl/M$ (the $\sl/M$ dependence is further constrained by (\ref{eq:EngC2})). This is quite unlike the weak gravity result $\sigma=M/4\pi\sl^2\sim M^{-1}$ for fixed $\sl/M$.  (\ref{eq:22scaling}) also implies that the ratio $A(r)/B(r)$ and the volume element factor $\sqrt{-g}$, as functions of $r/M$ inside the 2-2-hole, are independent of $M$.

The scaling behavior (\ref{eq:22scaling}) also determines the $M$ scaling of the coefficients in the series expansion of $A(r)$ and $B(r)$ such that $a_2, b_2\sim M^{-4}$ and $a_i\sim b_2b_i\sim M^{-i-2}$ for $i>2$. This then determines the leading in $1/M$ contributions to various quantities that are calculated in terms of these coefficients. We show the series expansions in this limit for generic and $\beta=0$ CQG in Appendix.~\ref{app:SeriesExp}. A few reference values are $[\sl/M,a_2M^4,b_2M^4,b_4M^2]=[0.6$, $0.053$, $6.0\times10^{-4}$, $0.48]$, $[1.8$, $0.104$, $1.1\times10^{-4}$, $0.34]$ for generic CQG and $[\sl/M$, $a_2M^4$, $b_2M^4]=[0.4$, $0.048$, $6.1\times 10^{-4}]$, $[1.8$, $0.009$, $1.4\times 10^{-4}]$ for $\beta=0$ CQG.

The surprisingly simple scaling behavior (\ref{eq:22scaling}) may be related to the dynamics around $r_H$, at which the drastic change of behavior of curvature invariants occurs. Outside $r_H$, curvature invariants follow the Schd prediction and are highly suppressed at large $M$. Around $r_H$, quadratic terms in the Lagrangian become comparable to the linear one, and curvature invariants start to respond to the dynamical scale $\Mp$. Inside $r_H$, since $A(r), B(r)\ll1$, curvature invariants with dimension $2n$ can be well approximated as $1/(A(r)r^2)^{n}$ times a function of the quantities $r^iA^{(i)}(r)/A(r), r^jB^{(j)}(r)/B(r)$.  The fact that curvature invariants around $r_H$ are mainly determined by the Planck scale dynamics suggests that $A(r_H)r_H^2$ should be quite independent of $r_H$. This provides some hint for the behavior of $A(r)$ and $I(r) $ in (\ref{eq:22scaling}). As for $B(r)$, although its overall scale is undetermined by field equations, its derivatives enter similarly to those of $A(r)$. 

Since we have a static configuration, we can consider a 4-volume that is a time interval $T$ times the integration over some spatial region $C$, i.e. $V\equiv \int_C drd\theta d\phi \sqrt{-g}$. Then (\ref{eq:22scaling}) implies that the interior region contribution to $V$ only grows with $M$. As an indication of how small the interior volume is, we find that a sphere centered at the origin having only one Planck volume has quite a large radius $r\sim 2.6M^{4/5}$. Another indication is that the proper distance from the origin out to a radius of $1.8M$ is only about one Planck length.\footnote{The fact that the ``radial size'' of a 2-2-hole is ${\cal O}(\LQG)$ brings a closer analogy to QCD where the size of hadronic states are characterized by the QCD scale.} We have mentioned that the vanishing volume is related to the finiteness of the action for the 2-2-hole. The scaling law implies that $S_{\mathrm{CQG}}\sim T M$ and we find that the coefficient is close to unity. 

\begin{figure}[!h]
  \centering%
{ \includegraphics[width=7.55cm]{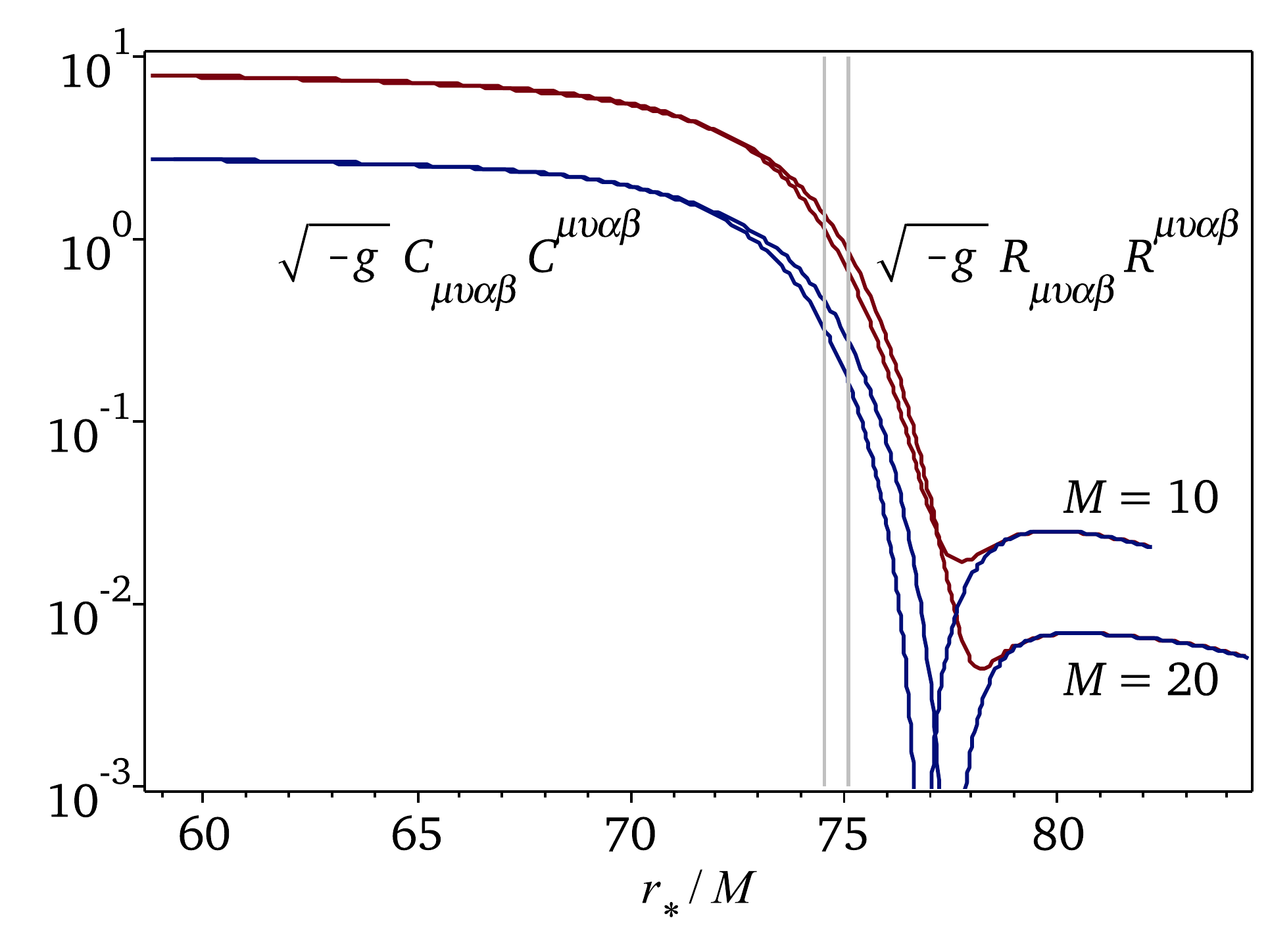}}\,\,\,
{ \includegraphics[width=7.45cm]{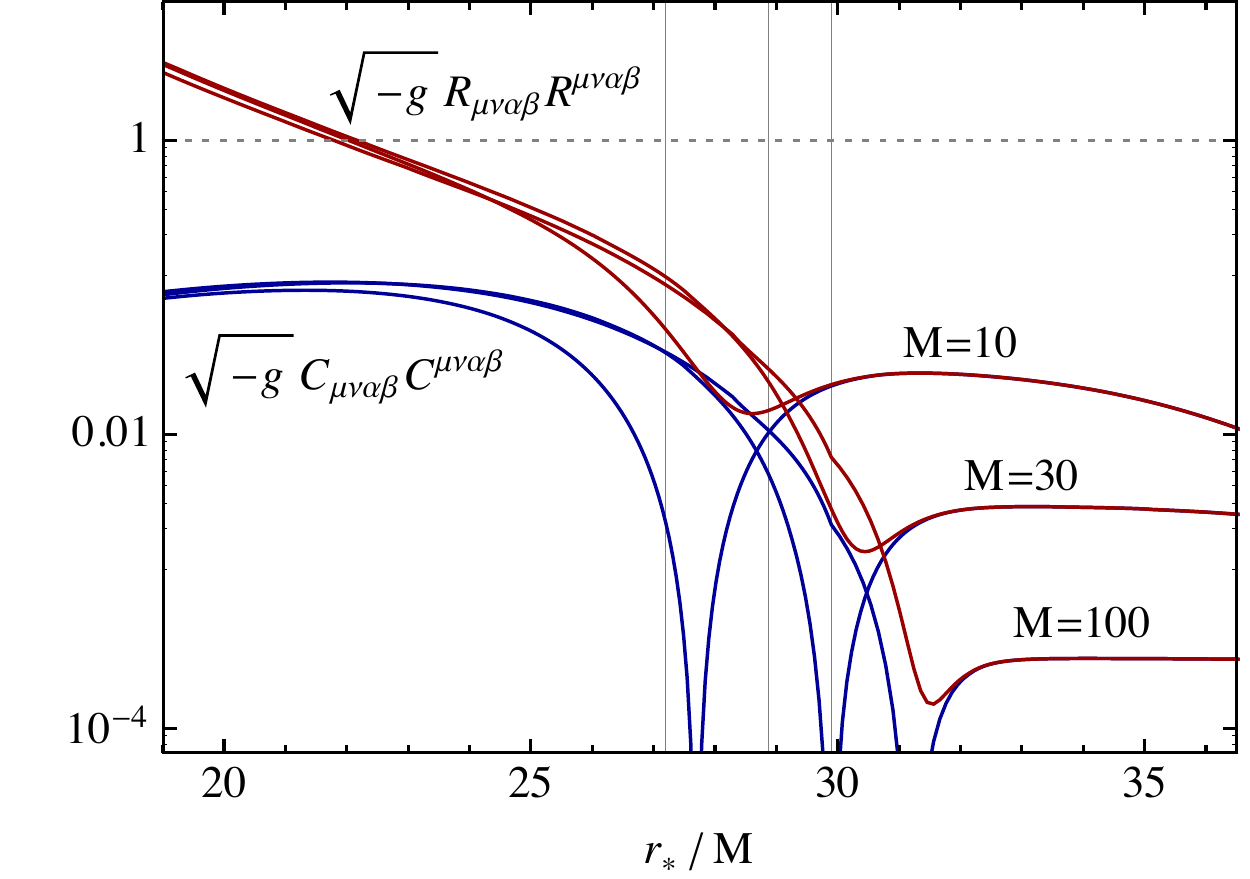}}
\caption{\label{fig:C2rstarM} 
$\sqrt{-g}C_{\mu\nu\alpha\beta}C^{\mu\nu\alpha\beta}$, $\sqrt{-g}R_{\mu\nu\alpha\beta}R^{\mu\nu\alpha\beta}$ as functions of $r_*/M$ for $\sl/M=1.8$ with different $M$ in generic CQG (left) and $\beta=0$ CQG (right). The vertical gray lines denote $r_H$ for each case.}
\end{figure}

Finally it is interesting to see how curvature invariants behave around $r_H$ and how they depend on $M$. We can zoom in on this region by using the tortoise coordinate $r_*$. Fig.~\ref{fig:C2rstarM} shows $\sqrt{-g}C_{\mu\nu\alpha\beta}C^{\mu\nu\alpha\beta}$, $\sqrt{-g}R_{\mu\nu\alpha\beta}R^{\mu\nu\alpha\beta}$ in generic CQG (left) and $\beta=0$ CQG (right).  At a particular radius outside of $r_H$ the Weyl tensor square drops zero. The two theories clearly differ in the interior and some of this difference is due to the different thin-shell models. At $r_H$ the curvature invariants are significantly below the Planck size for $\beta=0$ CQG, and the same is also true for generic CQG when the shell radius $\sl$ is smaller. In the full quantum theory, QQG, it could be that quantum effects only become very significant for a small radial range that is inside $r_H$ where the curvatures are not much smaller or larger than Planck size. At the location of the $A(r)/B(r)$ peak, quantum effects may still not be very significant. And for the rapidly growing curvatures in the deep interior the difference between the constant couplings of CQG and the running couplings of QQG should be of minor importance. Thus it could be that many of the properties of 2-2-holes that we have been discussing in CQG will continue to hold, in some approximation, in QQG.

\section{Physical properties of 2-2-holes}
\label{sec:phyProp}

The 2-2-hole may be the generic endpoint of gravitational collapse in quadratic gravity. In this section we explore some physical properties of 2-2-holes as the first step to relate them to astrophysical black hole candidates. In some of the discussion we shall be assuming that the thin-shell solutions generalize to solutions with more general matter distributions. One indication that this holds is the mild effect that a smooth matter distribution has on the series expansion of the (2,2) family. More details and other topics will have to be left for elsewhere.

\subsection{Radial stability}
\label{sec:radialSta}

Since we are studying the question of stability here we need to mention again that CQG has an intrinsic instability due to the presence of a Planck mass spin-2 ghost. But as we have discussed in Sec.~\ref{intro} we are assuming that this is not a feature of QQG in both the low and high curvature regimes. Thus we only use CQG to study stability with respect to the large scale perturbations in the matter distribution, which in our case we take to be a radial movement of the shell. 

As a common practice in GR the stability of a background solution against radial perturbations can be studied as a variational problem, and we shall carry over this procedure to CQG. For the thin-shell model the two field equations can be re-formulated as two equivalent equations. One specifies the physical mass as a function of the shell energy density and the shell radius, i.e. $M=M(\sigma,\sl)$. The other can be derived from the first variation of $M(\sigma, \sl)$ after implementing the conservation law (\ref{eq:EngC1}),
\begin{eqnarray}\label{eq:deltaM}
\delta M = \frac{\partial M}{\partial \sigma}\delta \sigma + \frac{\partial M}{\partial \sl}\delta \sl
=\delta \sigma\left(\frac{\partial M}{\partial \sigma}-\frac{\sl}{2(\sigma+p)}\frac{\partial M}{\partial \sl}\right)=0\,.
\end{eqnarray}
Then the radial stability can be inferred from the second variation of $M$, i.e. whether $\delta^2M>0$. It depends on the speed of sound $c_s^2\equiv\partial p/\partial \sigma$ of the shell matter. If $\delta^2M>0$ with $c_s^2\in(0,1)$ then a radially stable configuration can be supported by some reasonable matter. We implement the analysis for the $(0,0)$ and $(2,2)_E$ families respectively, each of which has a one-to-one mapping between $M$ and $\sigma$ for the respective range of $\sl$. 

Here we focus on the TS1 model in generic CQG. From (\ref{eq:deltaM}) the second variation of $M$ is
\begin{eqnarray}\label{eq:delta2M}
\delta^2M=\delta \sigma^2 \frac{\sl}{2(\sigma+p)^2}\frac{\partial M}{\partial \sl} c_s^2+...
= \frac{\delta \sigma^2}{\sigma+p}\frac{\partial M}{\partial \sigma} c_s^2+... \, ,
\end{eqnarray}   
where $...$ represents terms independent of the speed of sound. In the second step the coefficient of $c_s^2$ is simplified by (\ref{eq:deltaM}). With only numerical solutions we have no access to the full expression $M(\sigma, \sl)$, and the bound on $c_s^2$ cannot be derived analytically from $\delta^2M>0$. However we do know how $p$ and $\sigma$ change if the variation is restricted within the solution space for a given $M$. This defines the critical speed of sound $c_{s0}^2$, where $\delta^2M=0$ when $c_s^2=c_{s0}^2$,
\begin{eqnarray}\label{eq:Cspeedsound}
c_{s0}^2=\left.\frac{\partial p}{\partial \sigma}\right|_M
=\frac{dp/d\sl}{d\sigma/d\sl}
=\frac{p}{\sigma}+\frac{\sigma}{d\sigma/d\sl}\frac{d}{d\sl}\left(\frac{p}{\sigma}\right)\,.
\end{eqnarray}
$c_{s0}^2$ can be inferred from the $\sl$ dependence of $\sigma$ and $p/\sigma$ from numerical solutions with a given $M$ as in Fig.~\ref{fig:shellPM10}. Since $\sigma$, $p$ and $\partial M/ \partial \sigma$ are all positive, the radial stability condition $\delta^2M>0$ sets a lower bound on the speed of sound for the shell matter from (\ref{eq:delta2M}), i.e. $c_s^2>c_{s0}^2$. If $c_{s0}^2<1$ then it is possible that reasonable matter can support the 2-2-hole in a way that is stable against radial perturbations.    

\begin{figure}[!h]
  \centering%
{ \includegraphics[width=7.6cm]{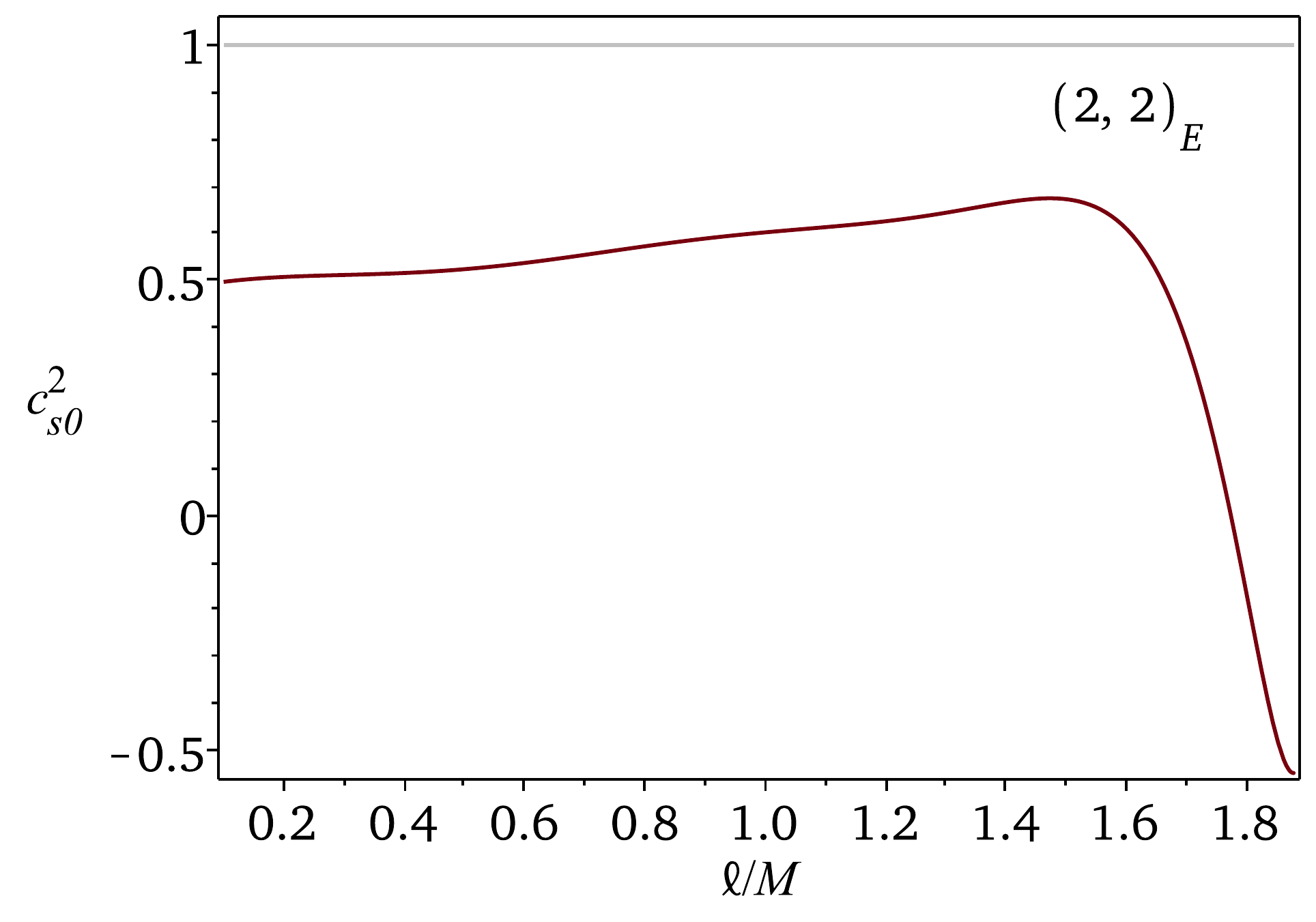}}\quad
{ \includegraphics[width=7.6cm]{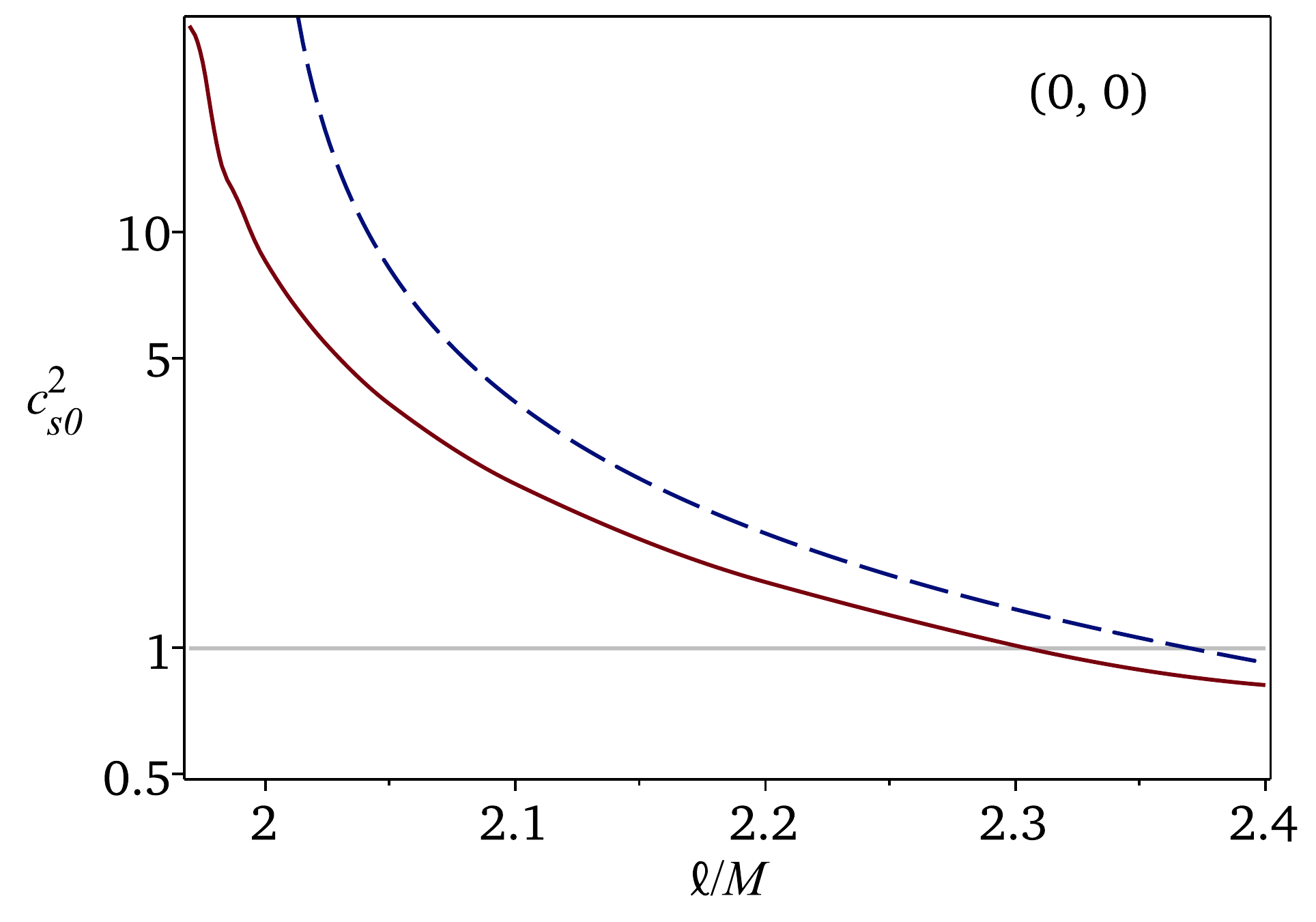}}
\caption{\label{fig:cs02}  The critical speed of sound at different shell radii with $M=10$ for the $(2,2)_E$ and $(0,0)$ solutions. The dashed line denotes the GR prediction.
}
\end{figure}

Fig.~\ref{fig:cs02} shows the critical speed of sound (\ref{eq:Cspeedsound}). For the $(2,2)_E$ case $c_{s0}^2<1$ holds at any $\sl$ that we can find a solution. Around the origin $c_{s0}^2\approx \sl B'(\sl)/4B(\sl) \approx 1/2$ is derived from the general form of the series expansion. The term with the gradient of $p/\sigma$ in (\ref{eq:Cspeedsound}) gives a negative contribution to $c_{s0}^2$ and causes $c_{s0}^2$ to drop down quickly at larger $\sl$. For large 2-2-holes the scaling law with respect to $M$ shows that $\sigma$ and $p/\sigma$ are only functions of $\sl/M$, and so the form of $c_{s0}^2$ in Fig.~\ref{fig:cs02} should apply for any $M$. The $(0,0)$ solution with $\sl$ not much larger than $r_H$ has $c_{s0}^2>1$, which implies radially instability for any reasonable matter. $c_{s0}^2$ decreases monotonically and approaches the GR prediction at large $\sl$. The condition $c_{s0}^2<1$ can be achieved at $\sl/M\gtrsim 2.3$.

Although we focus on the subclass $(2,2)_E$ in this work, the issue of stability raises the question of the existence of solutions in the $(2,2)$ family outside of $(2, 2)_E$. In the case of $\beta=0$ CQG we were able to find some examples of such solutions, where again $\sl\lesssim r_H$. So it may be the case that there is more than one solution for some given matter, each with a different $M$. Presumably only the one with the lowest $M$ can be stable. We leave a study of this extended solution space and its implications for stability for later.

The radial stability analysis in the thin-shell model provides some hint to how gravitational collapse proceeds in quadratic gravity. The $(0,0)$ solution with $\sl\gg r_H$ corresponds to a normal stable star. With more matter added onto the star an instability can develop, as we found for a shell when $\sl$ approaches $r_H$. A gravitation collapse occurs, concentrating matter further such that $(0,0)$ solutions no longer exist. With such dense matter and the existence of the horizonless $(2,2)_E$ solution, the object can turn into a 2-2-hole. Instead of metric components changing sign as for a black hole, a timelike singularity appears. And instead of matter moving inexorably towards the singularity of the black hole, an extended matter distribution can remain in the 2-2-hole. But while the 2-2-hole does seem to present a less pathological collapse scenario, certainly much more is needed to show that this is what actually occurs.

\subsection{Point particle geodesics and trapping }
\label{sec:geodesics}

Point particles geodesics provide the simplest way to probe a curved spacetime. 
On a static, spherically symmetric spacetime (\ref{ds2}) it suffices to study the geodesic  on the equatorial plane $\theta=\frac{\pi}{2}$. The motion is governed by two conservation laws,
\begin{eqnarray}\label{eq:geo0}
\frac{dt}{d\zeta}=\frac{E}{B(r)},\quad
\frac{d\phi}{d\zeta}=\frac{L}{r^2}\,.
\end{eqnarray}
The only nontrivial geodesic equation is for the radial motion,
\begin{eqnarray}\label{eq:geo1}
A(r)B(r)\left(\frac{dr}{d\zeta}\right)^2+B(r)\left(\frac{L^2}{r^2}+\vartheta\right)=E^2\,.
\end{eqnarray}
For massive (massless) particles $\vartheta=1\;(0)$, $\zeta$ is the proper time $\tau$ (the affine parameter $\zeta$). Since $A(r), B(r)$ remain regular and positive for the horizonless object, the qualitative features of the radial motion can be determined by the potential terms $B(r)(L^2/r^2+\vartheta)$ in (\ref{eq:geo1}). 

\begin{figure}[!h]
  \centering%
{ \includegraphics[width=7.8cm]{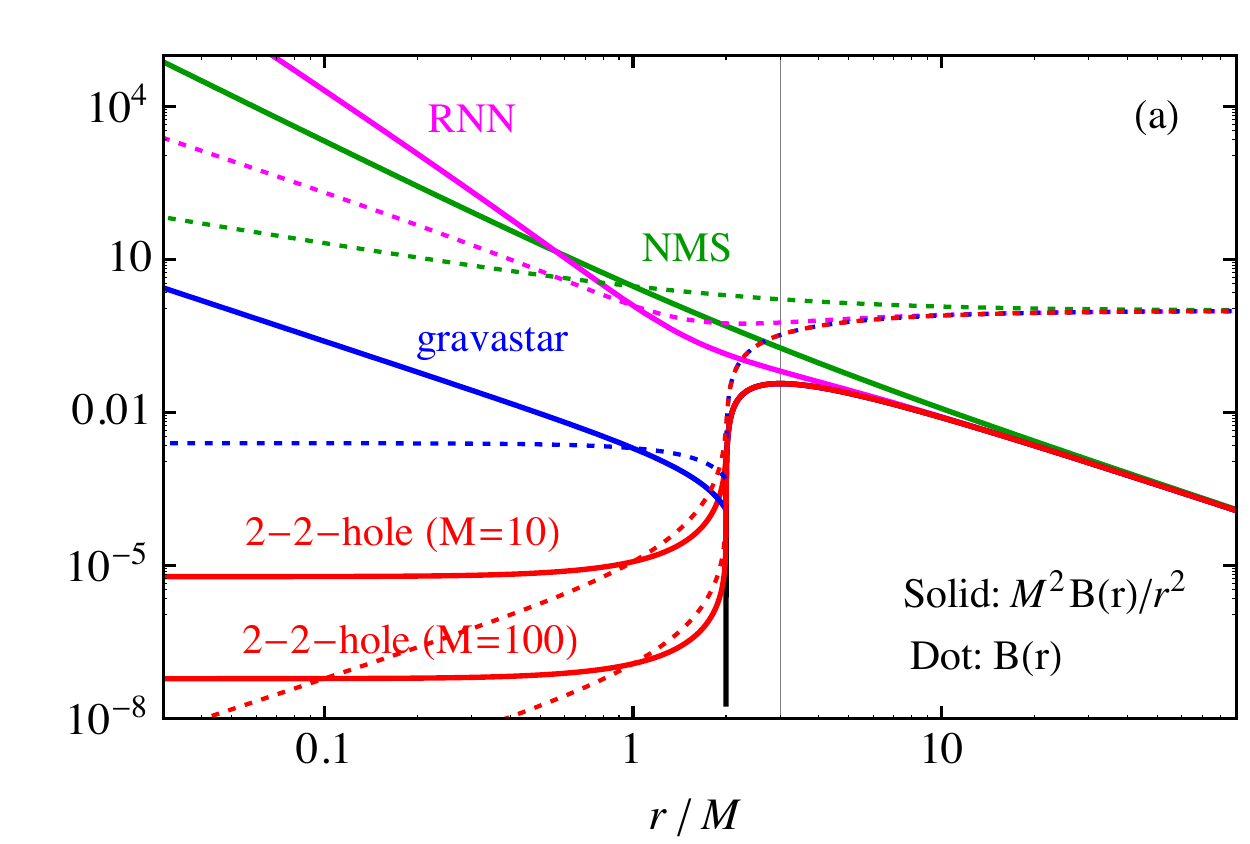}}\,\,
{ \includegraphics[width=7.2cm]{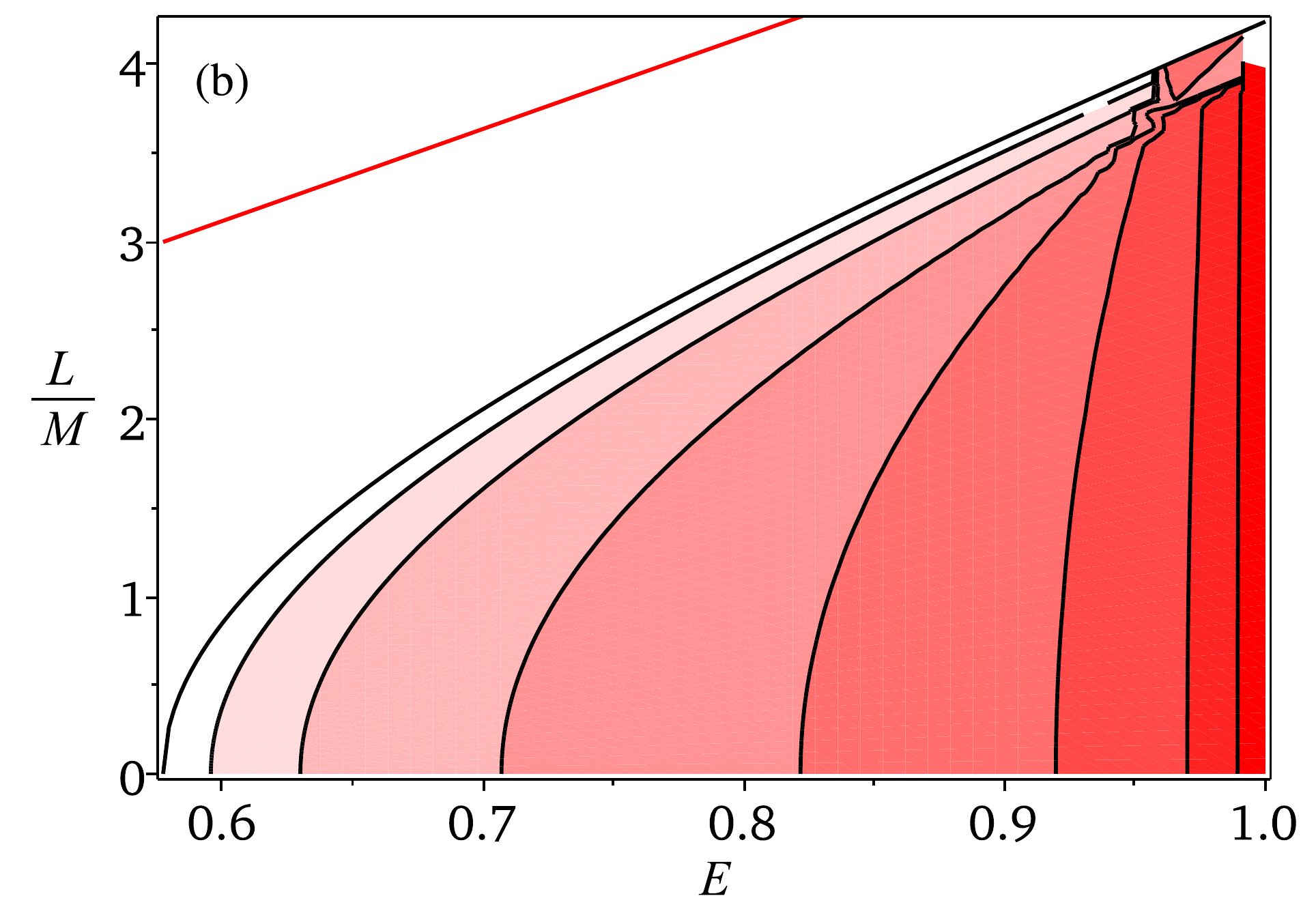}}\,\,
\caption{\label{fig:Vgeo} 
(a) The angular momentum potential $M^2 B(r)/r^2$ (solid) and the mass contribution $B(r)$ (dot) in the geodesic equation for different types of spacetime. (b) The values of $E$ and $L/M$ for which the turning point of massive particles is larger than $3M$. Black lines have turning points $3M$ and $(3+10^p)M$ for $p=-1,-.5,0,.5,1,1.5,2$ from left to right. Below the red straight line massless particles can escape to infinity.}
\end{figure}

We compare the $r$ dependence of the two terms in the potential, $B(r)/r^2$ and $B(r)$, in Fig.~\ref{fig:Vgeo}(a). The second term is only present for massive particles and it only dominates at large $r$. We compare different asymptotically-flat spacetimes, the negative mass Schd (NMS) spacetime, the Reissner-Nordström spacetime with a naked singularity (RNN) with $Q=1.5M$, and the gravastar with $R=2.001M$, $M_v=0.8M$.\footnote{The metric for the NMS: $B(r)=A(r)^{-1}=1+2M/r$ with $M>0$. The metric for the RNN: $B(r)=A(r)^{-1}=1-2M/r+Q^2/r^2$ with $Q>M>0$. The metric for the gravastar \cite{Visser:2003ge}: $B(r)=A(r)^{-1}=1-2M/r$ when $r>R$; $B(r)=cA(r)^{-1}=c(1-2M_vr^2/R^3)$ when $r\leq R$, with $c=(1-2M/R)/(1-2M_v/R)$ and $M>0$.} For the 2-2-hole we use $\beta=0$ CQG with $M=10, 100$ with $\sl/M=0.4$. The various potentials $B(r)/r^2$ differ drastically at small radius. For the RNN, NMS and gravastar spacetimes the diverging potential corresponds to the centrifugal repulsion for a particle with nonzero angular momentum. In contrast any geodesic that enters the 2-2-hole will go through the origin, as the potential inside $r_H$ drops down quickly and approaches a small constant at the origin, i.e. $B(r)/r^2|_{r=0}=b_2$. A 2-2-hole with astrophysical size is then characterized by an extremely deep gravitational potential since $b_2\sim 1/M^{4}$. A gravastar would need extreme fine tuning to achieve a potential this deep.

Circular orbits for massless particles, i.e. light rings, exist if $dr/d\zeta=0$ and $d^2r/d\zeta^2=0$, i.e. $2B(r_c)=r_c B'(r_c)$, $E^2/L^2=B(r_c)/r_c^2$.  The NMS and RNN spacetimes don't have any light rings. The Schd metric has an unstable light ring (the maximum of the potential) at $r_c/M=3$ with $M^2B(r_c)/r_c^2=1/27$, as denoted by the vertical line in Fig.~\ref{fig:Vgeo}(a). Massless particles with $L^2/E^2<27M^2$ can pass over the angular momentum barrier, both from the outside and the inside. As examples of ultra-compact objects, both the gravastar and the 2-2-hole closely resemble the Schd solution down to the would-be horizon and possess the same unstable light ring. The gravastar has another stable light ring at a smaller radius due to the potential having a minimum before diverging at the origin. This is a common feature of ultra-compact stars, which implies existence of long-lived modes and various types of instabilities \cite{Cardoso:2014sna}. Since $rB'(r)>2B(r)$ for $r\lesssim r_H$, the 2-2-hole has no additional light rings or circular orbits for massive particles in addition to what the Schd black hole has.

The geodesics inside a 2-2-hole are far from circular. From the two conservation laws the angular velocity of a geodesic in coordinate time is $d\phi/dt=L B(r)/E r^2$. But $L^2/E^2$ is bounded from above by $r^2/B(r)$. (This bound can be achieved for massless particles at a turning point, where $dr/d\zeta=0$ and $d^2r/d\zeta^2<0$.) Thus for geodesics  inside the 2-2-hole $d\phi/dt\lesssim \sqrt{b_2}\sim 1/M^{2}$. Meanwhile the radial velocity is $dr/dt\approx \sqrt{B(r)/A(r)}$ away from the turning point of the geodesic. Then for large $M$ the radial velocity is much larger than the angular velocity $rd\phi/dt$, and so away from the turning points the paths of geodesics in the interior are nearly straight lines (in this coordinate system).

What happens when particles of momentum $p_1^{\mu}$ and $p_2^{\mu}$ collide inside the 2-2-hole? The center of mass energy for a two particle collision is
\begin{eqnarray}\label{eq:Ecm}
\mathbb{E}_{\textrm{cm}}^2&=&g_{\mu\nu}(p_1^{\mu}+p_2^{\mu})(p_1^{\nu}+p_2^{\nu})
=m_1^2+m_2^2\\
&&+\frac{2}{B(r)}\left[\mathbb{E}_1\mathbb{E}_2-\kappa\sqrt{\mathbb{E}_1^2-B(r)\left(m_1^2+\frac{\mathbb{L}_1^2}{r^2}\right)}\sqrt{\mathbb{E}_2^2-B(r)\left(m_2^2+\frac{\mathbb{L}_2^2}{r^2}\right)}-\mathbb{L}_1\mathbb{L}_2\frac{B(r)}{r^2}\right]\nonumber
,\end{eqnarray}
where $p^{\mu}=m\, dx^{\mu}/d\tau$, $\mathbb{E}=m\,E$, $\mathbb{L}=m\, L$ for massive particles and $p^{\mu}=dx^{\mu}/d\zeta$, $\mathbb{E} =E$, $\mathbb{L}=L$ for massless particles. 
$\kappa=1, -1$ denotes whether the radial velocities of two particles are in the same or opposite direction. The center of mass energy can easily be enormous. For example a collision of radially moving particles with $\kappa=-1$ results in a center of mass energy $\mathbb{E}_{\textrm{cm}}^2\approx 4\mathbb{E}_1\mathbb{E}_2/B(r)$. This could be super-Planckian even at a radius $r$ that is not very close to the origin, due to the extreme smallness of $B(r)$ for a large 2-2-hole. Thus gravitation in the form of a 2-2-hole can yield a robust ultra-high energy particle collider.\footnote{There have been similar considerations for rapidly rotating Kerr black hole when collisions take place near the horizon with fine-tuned kinematics \cite{Harada:2014vka}. Collisions in horizonless spacetimes with regions of small $B(r)$ were considered in \cite{Patil:2012fu}.}

Supermassive particles and particles in hidden sectors can be created. While massless particles might escape to infinity (but see below), massive particles with $E<1$ cannot. Note that the enormous center of mass energies can result in large parton showers that dramatically increase the particle number and reduce the typical energy per unit mass $E$. This traps the massive particles by shifting the $E$ distribution lower. Particles on the high energy tail could still be observed in trapped noncircular orbits that extend outside of $r=r_H$ and even the light ring $r=\frac{3}{2}r_H$. The values of $E$ and $L/M$ for which the turning point is outside the light ring are shown in Fig.~\ref{fig:Vgeo}(b). 

The deep gravitational potential implies another phenomenon that shrouds this particle accelerator from outside observers even more. For example a particle with mostly radial momentum before the collision can be scattered to have mostly angular momentum after the collision. From (\ref{eq:Ecm}) and conservation of momentum, $L^2/E^2$ of the final particle can be as large as $\sim r^2/B(r)$. Since the latter scales like $b_2^{-1}\sim M^4$ and a massless particle needs $L^2/E^2<27M^2$ to escape, this particle now faces an enormous angular momentum barrier that prevents escape. The range of $L/M$ shown in Fig.~\ref{fig:Vgeo}(b) thus represents just a tiny fraction, of order $1/M$, of the range of possible $L/M$.

The escape probability can be estimated in the center-of-mass frame of the particle collision.
The propagation direction of a particle in the final state can be defined by the angle $\chi$ with respect to the radial direction in the orthonormal basis. Ignoring the particle's mass this is
\begin{eqnarray}
\tan\chi=\frac{r d\phi}{\sqrt{A(r)}dr}=\sqrt{\frac{L^2 B(r)/E^2r^2}{1-L^2B(r)/E^2r^2}}\,.
\end{eqnarray} 
The escape condition $L^2/E^2<27M^2$ then implies that $|\sin\chi|<3\sqrt{3B(r)}M/r$. The ordinary velocity vector must lie in a small cone of solid angle $\pi\chi^2$ around the radial direction. For an isotropic distribution of the final state particles, the portion of phase space where escape can occur is then tiny, $\chi^2\sim b_2M^2\sim1/M^2$.\footnote{Similar effects occur for the Schd metric. For example the escape cone for radiation from some surface located at radius $r=r_H+\sl_{\rm Pl}$ has a solid angle $\chi^2=27B(r_H+\sl_{\rm Pl})/4\sim1/M$ \cite{Abramowicz:2002vt}. Or consider radiation from an object located far from a black hole, $R\gg r_H$. Only a tiny portion of the radiation falls into the black hole because typically $L^2/E^2\sim R^2\gg27M^2$.} This shows an efficient trapping mechanism.  A particle falling into a 2-2-hole is easily pushed outside the escape cone through collisions with matter already in the interior, making it effectively trapped.

We can consider the effect of this trapping on the massless particles, since massive particles can be effectively trapped by the energy barrier. The time it takes for a significant fraction of the massless particles to escape is the cooling time. If there was no trapping of massless particles then this cooling time may be roughly the crossing time, that is of order $M$. The trapping means that only the massless particles in a tiny part of phase space at any given time can escape. This suggests that the actual cooling time is increased by a factor of order $M^2$. Thus a rough estimate of the cooling time of a 2-2-hole is $M^3/\Mp^4$ after re-introducing the factors of $\Mp$. This is an enormous time since it is of the same order as the lifetime of a similarly sized black hole.\footnote{Even if the use of the crossing time is not correct and the cooling time is a factor of $\Mp/M$ smaller, it is still many orders of magnitude larger than the age of the universe for $M\sim M_{\odot}$. Also the picture we have described here assumes that an interacting gas of particles is an appropriate description of matter inside the 2-2-hole.} It also implies that the luminosity of a 2-2-hole is extremely small.

During a gravitational collapse a 2-2-hole may be formed when enough mass falls inside a would-be horizon. Then internal collisions populates a large trapped phase space as we have described. A tiny fraction of the particles can exist on orbits that temporarily escape the 2-2-hole and fall back in, to form a trapped cloud. But the cloud of massive particles that could exist outside the light ring should disappear due to degradation of typical $E$'s from inelastic collisions. Any later accretion of matter onto the 2-2-hole would seem to be effectively absorbed with very little induced emission due to the trapping mechanism. Accretion would simply cause an increase in the 2-2-hole mass. If so then the 2-2-hole is more like a black hole in this respect and it may escape constraints on surface emission \cite{Abramowicz:2002vt,Broderick:2009ph}. 

\subsection{Regular field dynamics}
\label{sec:singularity}

We have seen that interior geodesics end at the origin within a finite proper time. The geodesic incompleteness is a common way to define a singularity and it plays an important role in the proof of the singularity theorem in GR.  But this may or may not point to an actual physical ambiguity. Due to the asymptotically free nature of QQG, probing the timelike singularity in the high curvature region should be addressed within quantum field theory (QFT) in curved spacetime. The probes are then the particle states of the QFT. These states include the graviton as well as particles in the matter sector that may well also be asymptotically free. But before dealing with QFT we should first confirm that relativistic classical field theory is well defined. And within classical field theory we can consider finite energy wave packets as the probes of interest. Indeed, the non-relativistic Schrodinger equation seems not to be very useful here since, as we have seen, particles tend to be highly relativistic around the singularity.

Here we consider the Klein-Gordon equation for the massless spin-0 scalar field. With the line element (\ref{ds2}),  $\Box\varphi=0$ becomes
\begin{eqnarray}\label{eq:KGE1}
\partial _t^2 \psi_l =\frac{B}{A}\partial _r^2\psi _l+\frac{B}{A}\left(\frac{2}{r}+\frac{B'}{2B}-\frac{A'}{2A}\right)\partial _r\psi _l- B\frac{l (l+1)}{r^2}\psi _l\equiv \mathbb{A}\psi_l\,.
\end{eqnarray}
The angular variables are separated using spherical harmonics $\varphi=\sum_{lm}\psi_l(r,t)Y_{lm}(\theta,\phi)$. The spacetime could be defined with the singular point at $r=0$ removed. Approaching the problem naively, one could still examine the behavior of solutions of (\ref{eq:KGE1}) around the origin. Any solution that has diverging energy as $r\to 0$ could be discarded on physical grounds. If for each $l$ there is a unique remaining solution then it would appear that the evolution of classical fields proceeds without ambiguity.

We can put this in the context of an existing mathematical procedure for defining the field dynamics on a singular and so-called inextendible spacetime, as introduced by Wald \cite{Wald:1980jn}. Here $\mathbb{A}$ is viewed as an operator on a Hilbert space of fields on a constant time hypersurface $\Sigma$. The problem is to see whether there is a unique positive self-adjoint extension of the operator $\mathbb{A}$ on the Hilbert space, as denoted by $\mathbb{A}_E$. This ``essentially self-adjoint'' operator \cite{Horowitz:1995gi}, if it exists, generates a solution from the initial data via a time translation using $\mathbb{A}_E^{1/2}$.
The initial value problem of the wave equation (\ref{eq:KGE1}) is then well-posed and the singularity has introduced no ambiguity.

The existence of $\mathbb{A}_E$ is tied to the appropriate choice of the Hilbert space. We follow \cite{Ishibashi:1999vw} to define the Hilbert space as the first Sobolev space $\mathcal{H}^1$. This requires that both the field and its first derivative be square integrable. In particular the Sobolev norm is chosen such that its finiteness is equivalent to the finiteness of the energy $E=\int_\Sigma d \Sigma \,n_\mu T^{\mu\nu}\xi_\nu$.\footnote{$n_\mu=\sqrt{B}\delta^t_\mu$ is the unit normal vector to $\Sigma$. The induced metric on $\Sigma$ is $h_{ij}=\mathrm{diag}(A(r),\, r^2,\, r^2\sin\theta^2)$ and $d\Sigma=\sqrt{h}drd\theta d\phi$.} This energy is conserved for a static background due to the existence of the timelike Killing vector $\xi_\nu$ and the fact that $T^{\mu\nu}\xi_\nu$ is a conserved current. Thus the Hilbert space of finite energy configurations is consistent with the time evolution. In fact a conserved energy was used in \cite{Ishibashi:2003jd} to prove that the mathematical procedure in \cite{Wald:1980jn} represents the only possible way to define the dynamics of a scalar field in a static, non-globally-hyperbolic spacetime.

The Sobolev norm can be chosen to be   
\begin{eqnarray}\label{eq:Sobnorm}
||f||^2=\frac{1}{2}\int_{\Sigma} d\Sigma \, B^{-1/2} f^* f + \frac{1}{2} \int_{\Sigma}	d\Sigma \, B^{1/2}h^{ij}D_i f^* D_j f
.\end{eqnarray}
For the test scalar field, after separation of angular variables, this becomes
\begin{eqnarray}\label{eq:Sobnorm2}
||\varphi||^2=\sum_{l} \int_{0}^\infty dr r^2\left[\frac{1}{2}\sqrt\frac{A}{B} \psi_l^2 + \frac{1}{2} \sqrt\frac{B}{A}   \left(\psi'_l\right)^{2} \right]\,.
\end{eqnarray} 
The essential self-adjointness is guaranteed if only one solution of $\mathbb{A}=0$ as defined in (\ref{eq:KGE1}) has finite Sobolev norm in the small $r$ region.\footnote{The mathematical procedure is to study $\mathbb{A}\psi_l=\pm i\psi_l$, but the $i\psi_l$ term will be irrelevant at small $r$. The same is true for a mass term.} We compare the 2-2-hole with the gravastar and the singular NMS and RNN spacetimes around the origin in Tab.~\ref{tab:scalarfieldeq}. This shows the behavior of two linearly independent solutions at small $r$, and the number of solutions with finite norm in the final column. For the first three spacetimes the second solution $\psi_{l2}(r,t)$ is not Sobolev finite. Thus for these spacetimes $\mathbb{A}_E$ exists and the timelike singularity is regular as probed by finite energy wave packets.
\begin{table}[h]
\begin{center}
\caption{Near origin behaviors for different spacetimes}
\vspace{1em}
\begin{tabular}{|c|c|c|c|c|c|}
\hline
&&&&&
\\[-3mm]
Spacetime\,\, & $A(r)$ & $B(r)$ & $\psi_{l1}(r,t)$ & $\psi_{l2}(r,t)$ & Num
\\
&&&&&
\\[-3.5mm]
\hline
2-2-hole & $r^2$ & $r^2$  & 1 & $r^{-1}$ & 1
\\
\hline
gravastar & $r^0$ & $r^0$ & $r^{l}$ & $r^{-(l+1)}$  &  1
\\
\hline
NMS & $r$  &  $r^{-1}$ & 1 & $\ln r$ & 1
\\
\hline
RNN & $r^2$  &  $r^{-2}$ & 1 & $r $ & 2
\\
\hline
\end{tabular}
\label{tab:scalarfieldeq}
\end{center}
\end{table}

For the 2-2-hole we see that the small $r$ behavior of waves is independent of angular momentum; indeed they all behave like the $S$-wave on a nonsingular spacetime. We have already seen that geodesics do not see an angular momentum barrier at the origin. Also, only for the 2-2-hole, $\psi_{\sl 1}(r,t)$ actually satisfies a Neumann boundary condition at $r=0$, namely $\left.\partial\psi_{\sl 1}(r,t)/\partial r\right|_{r=0}=0$ for any $\sl$.

For the RNN spacetime the operator $\mathbb{A}_E$ does not exist. The problem is also apparent by seeing that the allowed solutions in this case can imply a loss of unitarity (a net flux in or out of the singularity). To obtain a sensible boundary condition for the RNN spacetime one might impose unitarity as an external constraint \cite{Chirenti:2012fr}. This is not necessary for the other spacetimes.  

We note that both the NMS and RNN spacetimes are exact vacuum solutions of CQG. But like the Schd solution, none of these exact solutions are sourced by matter. Since the theory provides another set of solutions, the (2, 2) solutions, which along with the (0, 0) solutions shows how spacetime actually responds to matter, these exact singular solutions are relegated to providing useful approximations in vacuum regions where the CQG corrections are exponentially small. (The NMS solution may not even have this role to play or otherwise there would presumably be a vacuum instability \cite{Horowitz:1995ta}.)

We only briefly speculate about how states as described by QFT would interact with the 2-2-hole background. The problem here is somewhat analogous to the treatment of Rutherford scattering as an external field problem in QFT. But there are complications. We have seen that particles can be accelerated to Planckian energies as they fall in. Thus when there is a momentum transfer between the background field and the particle, this momentum may also be Planckian in size.  But with Planckian momentum transfers the effective graviton-matter coupling is of order one. Thus rather than simply scattering, the particle may initiate a type of graviton parton shower in the 2-2-hole interior.

\subsection{A brick wall and entropy}
\label{sec:brick}

Here we consider the statistical mechanics of a quantized scalar field in the background of a 2-2-hole. It proves to be very simple to carry over the analysis initiated in \cite{tHooft:1984kcu} and further interpreted in \cite{Mukohyama:1998rf}, for the brick wall model of the black hole entropy. In this model the scalar wave equation on the Schd background is considered with a Dirichlet boundary condition at the ``brick wall'' located just slightly outside the black hole horizon. This gives a discrete set of modes when the fields are also required to vanish at a large radius $L\gg r_H$. The 2-2-hole gives a very similar problem where we can return to the scalar equation (\ref{eq:KGE1}) and use the Neumann boundary condition $\psi_\sl'(0)=0$ that we have already motivated. Now the ``brick wall'' is at the origin\footnote{In both problems the wall is at a finite value of the tortoise coordinate $r_*$.} and the result is again a discrete set of modes. The WKB approximation used in \cite{tHooft:1984kcu,Mukohyama:1998rf} can also be carried over. Since the analysis remains so similar we just give the results. The entropy $S$ and total thermal energy $U$ are
\begin{align}
&S=\frac{(2\pi)^3}{45}\int_0^L T(r)^3A(r)^{1/2}r^2dr,\\
&U=\frac{3}{4}T_\infty S.
\end{align}
$T(r)\equiv T_\infty/\sqrt{B(r)}$ is the local temperature. The large volume contribution to $S$ that scales like $L^3$ is not of interest here. Compared to the original results, the inverse relationship between $A(r)$ and $B(r)$ that is assumed in \cite{tHooft:1984kcu,Mukohyama:1998rf} has been relaxed and the $r$ integral now ranges down to zero.

The observation from \cite{Mukohyama:1998rf} is that there is another contribution to the local energy density that should appear in the field equations, and that is the background dependent renormalized vacuum energy density. Since the metric is horizonless and static, the fields are naturally quantized with respect to the Killing time. This leads to the negative Boulware vacuum energy density. For an appropriate $T_\infty$ and in the region where $T(r)$ is reaching its highest values, the two contributions to the energy density can be made to cancel. In the case of \cite{Mukohyama:1998rf} this occurs when $T_\infty=T_{\rm Hawking}\equiv 1/8\pi M$. Then for this temperature the sum of these two energy densities causes negligible or minor back-reaction on the metric, and in this way the temperature is determined in a self-consistent way.

In the respective calculations of the thermal and vacuum energy densities, very similar mode sums are being performed. Their cancellation in the region where each are receiving large small wavelength contributions indicates that the effective UV cutoff $T(r)$ in the thermal case is being chosen to match the effective UV cutoff in the renormalized vacuum energy calculation. We expect that the same arrangement can be made for the 2-2-hole background for some $T_\infty\propto T_{\rm Hawking}$. For the 2-2-hole a third contribution to the energy density is the nonthermal matter component, as represented in our solutions by the thin shell of matter. When the vacuum and thermal components largely cancel, then we can expect the metric solution to be very similar to what we already have. But even if they don't largely cancel there may still be other, less similar 2-2-hole solutions. 

Let us focus on the $M$ dependence of the entropy in the case $T_\infty\propto T_{\rm Hawking}$. We have
\begin{align}
S=\left(\frac{T_\infty}{T_{\rm Hawking}}\right)^3\frac{1}{2880}\frac{1}{M^3}\int_0^L A(r)^{1/2}B(r)^{-3/2}r^2dr.
\label{SS}\end{align}
Besides the trivial $L^3$ contribution, the dominant contribution comes from the interior of the 2-2-hole where the large $M$ scaling law for $A(r)$ and $B(r)$ applies. The integrand in (\ref{SS}) is quite uniform in $r$ in the interior and it give a contribution to the integral that scales like $M^5$. This results in an area law for the entropy of a 2-2-hole, $S\propto M^2$.

This entropy due to a single scalar field with $T_\infty\approx T_{\rm Hawking}$ turns out to be similar in size to the Bekenstein-Hawking entropy. In the original brick wall model the wall location has to be tuned to obtain such a value. Our numerical solutions also show that the contribution to $S$ from the region $r>r_H$ where $A(r),B(r)$ still differ significantly from unity is 3 or 4 orders of magnitude smaller. And this contribution is even less significant in our extrapolation to very large 2-2-holes as long as the power $\eta$ discussed in the next section is not greater than 2.

Thus we have found an intriguing connection between the scaling behavior of the interior 2-2-hole solution and an area law for entropy. Also for $T_\infty\approx T_{\rm Hawking}$ the total thermal energy $U$ is of order $M$. The local temperature of this thermal component in the deep interior is super-Planckian, $T(r)\gtrsim M/r$, and so the timelike singularity is effectively shrouded by its own fireball.

\subsection{A time delay to probe the internal structure}
\label{sec:QNMs}

Recently it is argued that whether an horizonless ultra-compact object or a black hole is formed from a compact binary coalescence, the early stage of the post-merger ringdown phase of the gravitational waves can be identical \cite{Cardoso:2016rao} \cite{Cardoso:2016oxy}. This was seen by studying some toy models of metric and scalar perturbations for various ultra-compact objects. The initial ringdown waveform is associated with the excitation of the unstable light ring. But unlike a black hole, the resulting wave that enters the compact object can be reflected back by the interior. We have seen that the 2-2-hole has such a reflecting type of boundary condition at $r=0$. The wave coming back out can then be partially reflected back in due to the light ring barrier. The result is a series of echoes of the initial ringdown. In principle some information about the oscillation modes of the interior can be imprinted on these echoes. But the numerical examples in \cite{Cardoso:2016oxy} seem to suggest that at least for the first few echoes, the light ring modes are simply being re-excited after each round trip of the interior travelling wave. It is the time delay between echoes that may provide the most accessible information about the interior. Our main interest here then is to estimate what this time delay is for an astrophysical 2-2-hole. We shall also look at some more features of the wave equation that are peculiar to the 2-2-hole.

The time delay is roughly the coordinate time that light takes to traverse through the ultra-compact object starting from the light-ring radius of $r=3M$. Since the radial speed of light in coordinate time is $\sqrt{B(r)/A(r)}$, the time delay is 
\begin{eqnarray}
\Delta t =2\int _0^{3M} \sqrt\frac{A(r)}{B(r)}dr= 2\int _{r_*(0)}^{r_*(3M)} dr_*\,.
\end{eqnarray}
In other words it is twice the range of the tortoise coordinate from the origin to the light ring. For the 2-2-hole with large $M$, the ratio $A(r)/B(r)$ reaches a narrow peak around $r_H$ before sharply dropping to a constant in the interior. We find that $\Delta t$ is quite sensitive to the behavior of solutions around the peak region. To explore this we use the thin-shell solutions found in $\beta=0$ CQG with $M\in[10,10^4]$ and $\sl/M=0.4$ for the extrapolation to an astrophysical 2-2-hole.

\begin{figure}[!h]
  \centering%
{ \includegraphics[width=7.7cm]{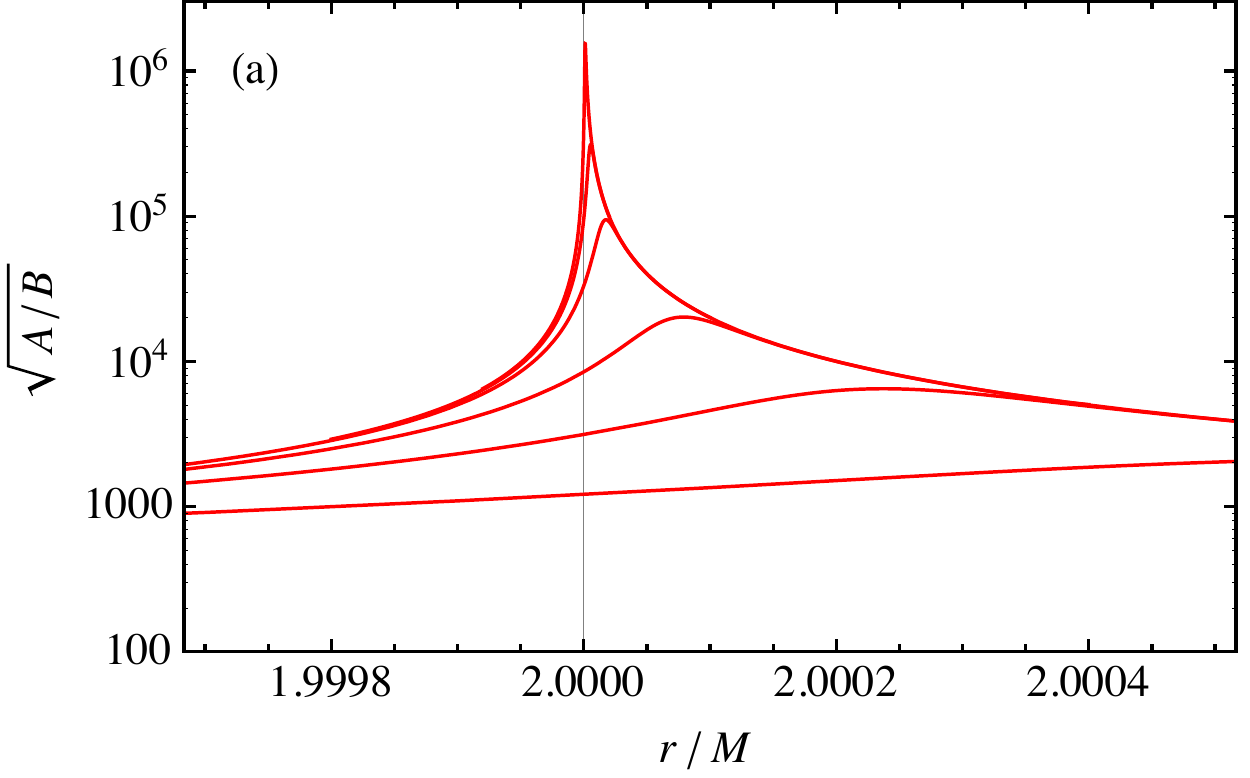}}\,\,
{ \includegraphics[width=7.2cm]{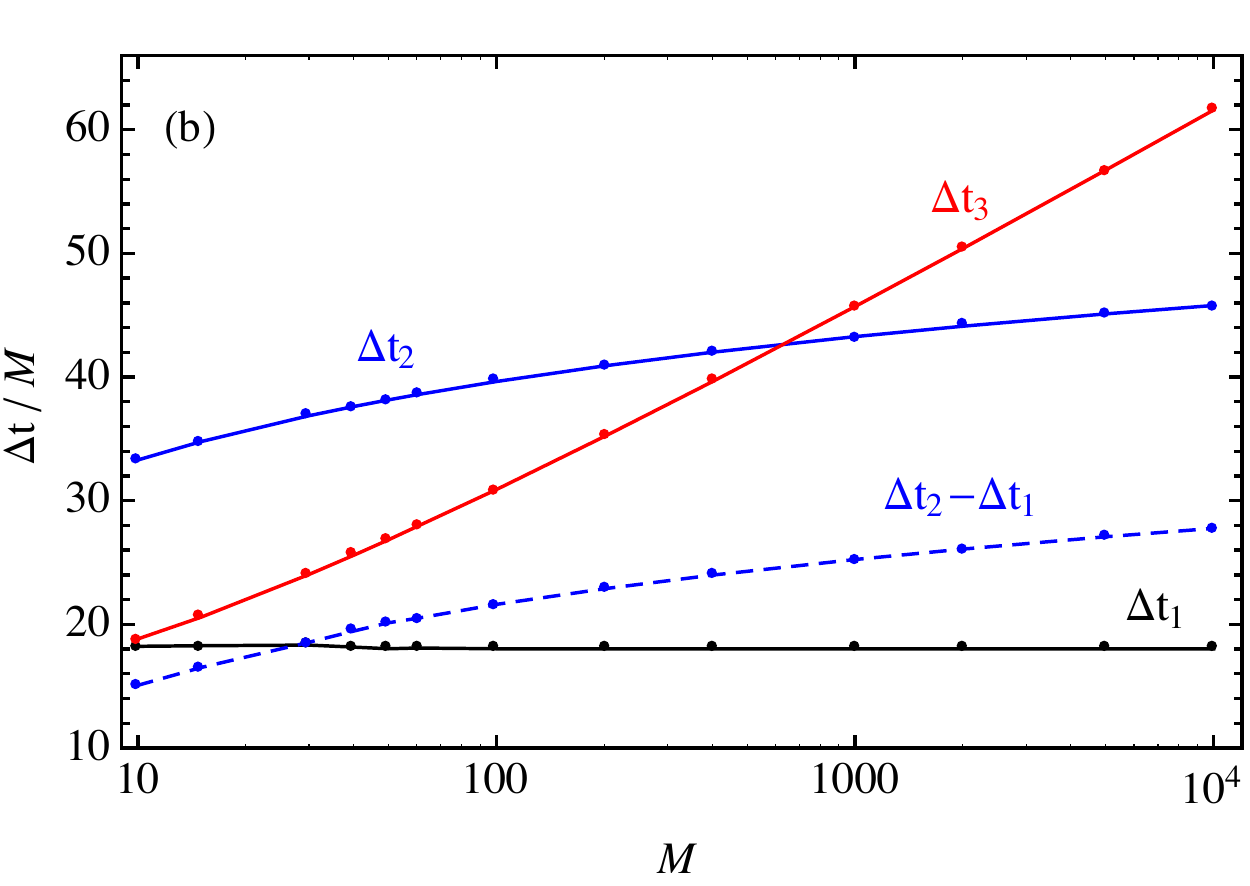}}
\caption{\label{fig:timedelay} 
(a): $\sqrt{A(r)/B(r)}$ around $r_H/M=2$ for $M=100$, 200, 400,1000, 2000, 5000 with $\sl/M=0.4$ from bottom to top. (b): $\Delta t/M$ as function of $M$, with $\Delta t_1$ (black), $\Delta t_2$ (blue), $\Delta t_3$ (red) for the integration within $[0,M]$, $[M,r_\mathrm{peak}]$, $[r_\mathrm{peak},3M]$ respectively  (the blue dashed line denotes $\Delta t_2-\Delta t_1$).}
\end{figure}

Fig.~\ref{fig:timedelay}(a) shows the growth of the peak of $\sqrt{A(r)/B(r)}$ near $r_H$ for increasing $M$. Assuming the peak value occurs at a radius $r_{\mathrm{peak}}=r_H(1+\delta)$ that is also close to where the deviation from the Schd metric occurs, then $\sqrt{A/B}_{\mathrm{peak}}\sim \left(1-r_H/r_{\mathrm{peak}}\right)^{-1}\approx 1/\delta$. Our numerical results are used to determine the power $\eta$ in $\delta\sim1/M^\eta$. We find $\eta$ to be slowly increasing up to $M\sim10^4$, where $\eta\approx 1.75$, and that it appears consistent to have an asymptotic value around 2 (or at least to be around 2 for an astrophysical sized $M$). The proper distance from the peak to the would-be horizon is roughly $\sqrt{A}_{\mathrm{peak}}(r_{\mathrm{peak}}-r_H)\sim r_H\sqrt{\delta}\sim M^{1-\eta/2}$. Thus our estimate with $\eta\approx2$ turns out to be consistent with the proper distance being of order the Planck length.

To see how $\Delta t$ varies with $M$, we split the integration into three regions: $\Delta t_1$ for $[0,M]$, $\Delta t_2$ for $[M,r_{\mathrm{peak}}]$ and $\Delta t_3$ for $[r_{\mathrm{peak}},3M]$. The outer contribution $\Delta t_3$ turns out to be the largest and it is closely approximated by an integration of $1-2M/r$ from $r_{\mathrm{peak}}$ to $3M$ which gives $\Delta t_3/M\sim 4\ln \delta^{-1}\sim 4\eta \ln M$. For our extrapolation of  $\Delta t_3$ we use $\eta=1.75$ and 2 to set the lower and upper bounds respectively.  The integration from $[0,M]$ is roughly a constant, with $\Delta t_1/M\sim 2\sqrt{a_2/b_2}$, whereas the integration from $[M,r_{\mathrm{peak}}]$ also includes the inner part of the peak. The latter can be defined by $\Delta t_2 - \Delta t_1$ and is found to increase gradually. Fig.~\ref{fig:timedelay}(b) shows that $\Delta t_2 - \Delta t_1$ grows slower than $\Delta t_3$, which can also be seen from the asymmetric shape of the peak in Fig.~\ref{fig:timedelay}(a). We assume that the extrapolation of $\Delta t_2 - \Delta t_1$ is bounded by two straight lines as functions of $\ln M$, one with zero slope and the other with the slope of $\Delta t_2 - \Delta t_1$ at $M=10^4$. Combining the various contributions our estimate for the time delay is
\begin{eqnarray}
700+7\ln\frac{M}{30M_{\odot}} \lesssim \frac{\Delta t}{M}
\lesssim 
860+9\ln\frac{M}{30M_{\odot}}\,.
\end{eqnarray}
For $M\sim 30\,M_{\odot}$ the time decay $\Delta t$ is in the 100-125 ms range. It should be noted that this estimate is obtained for a particular $\sl/M$ within $\beta=0$ CQG. But it indicates that an astrophysical 2-2-hole naturally has a time delay $\Delta t$ significantly longer than the black hole ringdown damping time of a few ms. An initial analysis of Advanced LIGO data has already looked for echoes of the ringdown with time delays that are similar to our estimate \cite{Abedi:2016hgu}.

\begin{figure}[!h]
  \centering%
{ \includegraphics[width=7.35cm]{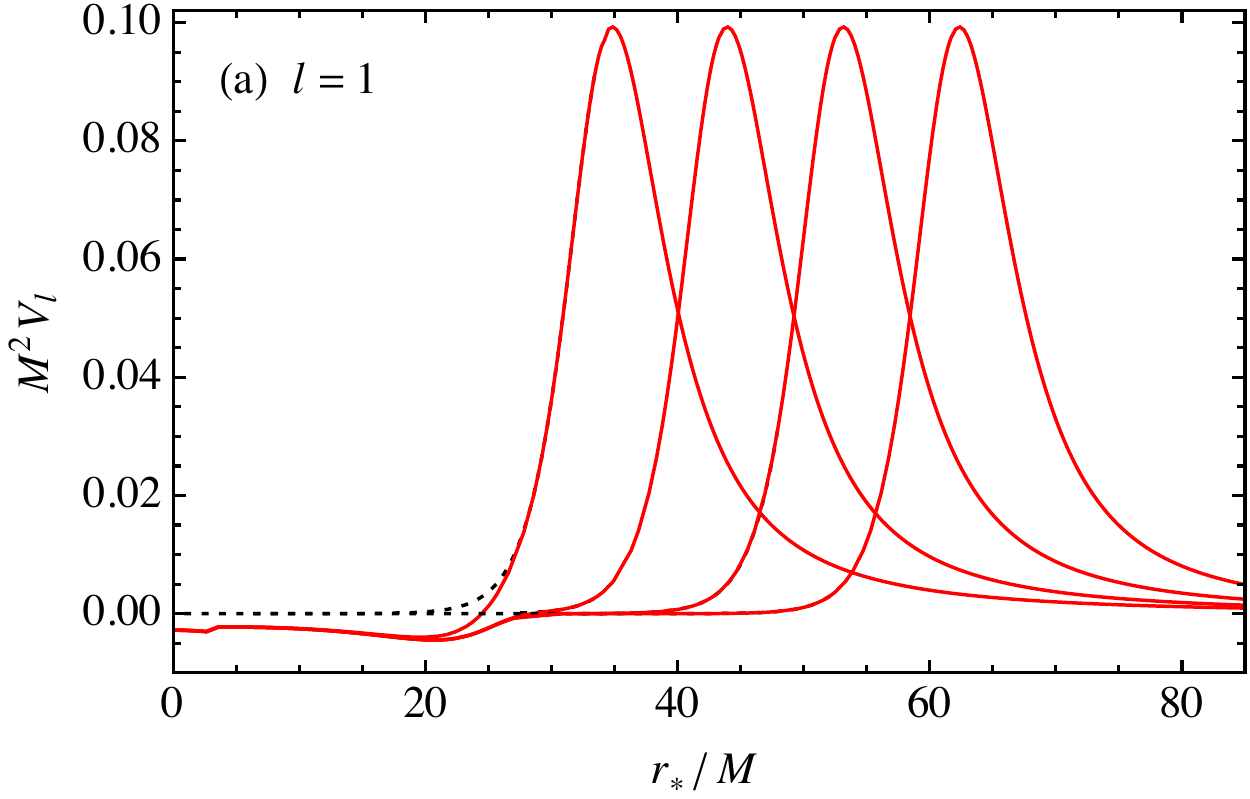}}\,\,
{ \includegraphics[width=7.75cm]{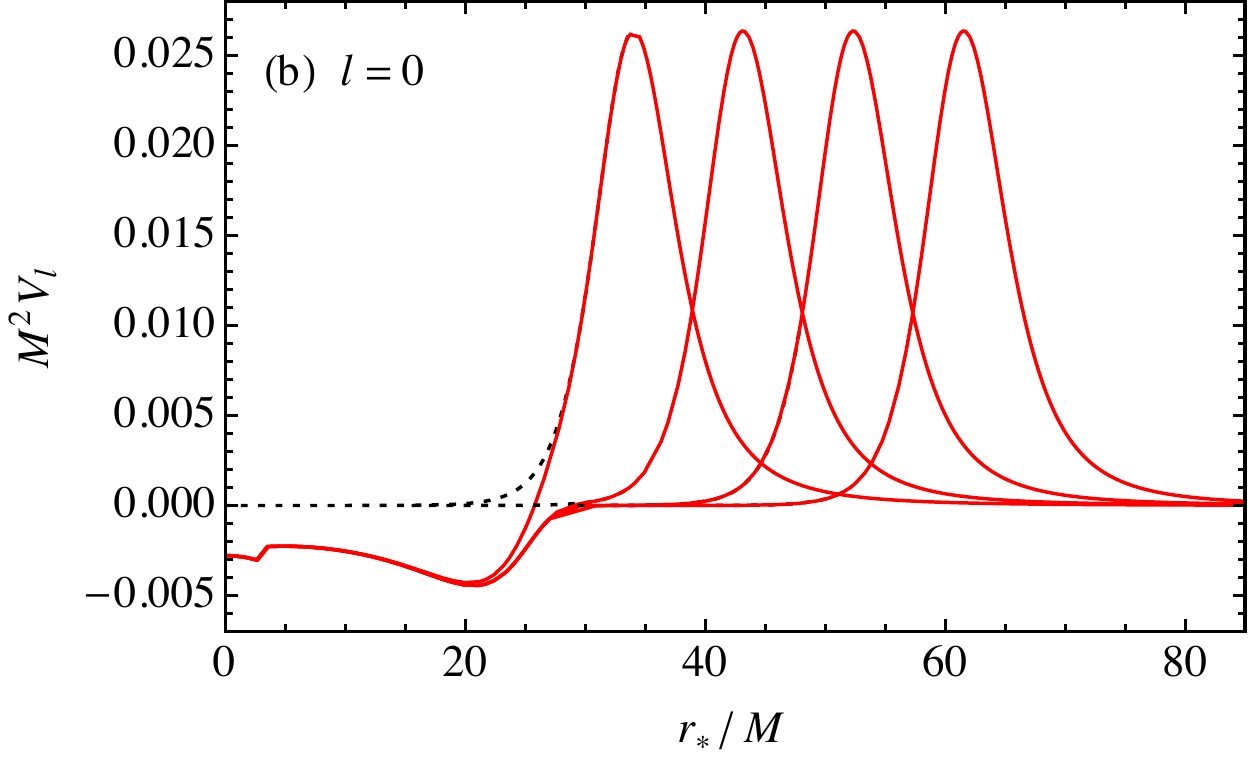}}
\caption{\label{fig:Vl22Schd solution} 
The potential $M^2V_l(r)$ with (a) $l=1$ and (b) $l=0$ as a function of $r_*/M$ with $M=10$, $100$, $1000$, $10^4$ from left to right and $\sl/M=0.4$ in $\beta=0$ CQG. The black dotted lines denote the Schd solution with $r_*$ shifted to match the peaks.}
\end{figure}

Next we explore more features of the wave equation. It is convenient to rewrite the radial equation  (\ref{eq:KGE1}) to make it resemble the Schrodinger equation. The linear derivative term is eliminated by defining $\psi_l(r,t)=e^{-i\omega t}\Psi_l(r)/r $ and by using the tortoise coordinate,  
\begin{eqnarray}\label{eq:mastereq}
(\partial _{r_*}^2+\omega^2-V_l(r))\Psi _l=0,\quad
V_l(r)=B(r)\frac{l (l+1)}{r^2}+\frac{1}{2r}\frac{B(r)}{A(r)}\left(\frac{B'(r)}{B(r)}-\frac{A'(r)}{A(r)}\right)\,.
\end{eqnarray}
For the 2-2-hole and the black hole, $V_l(r)$ is the same until $r$ is very close to $r_H$. For the black hole, $V_l(r_H)=0$ where $r_H$ corresponds to $r_*=-\infty$. For the 2-2-hole $r$ extends down to $r=0$ which now corresponds to a finite value of $r_*$ and where $V_l(0)$ is finite. We can then choose $r_*(0) = 0$. For the NMS and RNN spacetimes, $V_l(r)$ in contrast has a singular $1/r_*^2$ behavior.

In Fig.~\ref{fig:Vl22Schd solution} we display the potential $M^2V_l(r)$ of the 2-2-hole, for various $M$ and for $l=1$ and 0. When $l\neq0$ the potential at large radius is dominated by the unstable light ring peak at $r=3M$ with $M^2V_l(r)\approx l^2/27$. We see that the main effect of increasing $M$ is just to shift the peak position to larger $r_*$. This corresponds to the increase in $\Delta t/M$ that we have just discussed. The contribution from $\Delta t_3$ to the shift in the peak occurs in the region where there is still agreement with the Schd solution. This is becoming more evident for the larger values of $M$ in agreement with Fig.~\ref{fig:timedelay}(b).

At small $r$, $V_l(r)$ becomes independent of $l$ and so we see again that waves of different $l$ behave like S-waves in the interior. In fact in the interior region the $l$-independent term in $V_l(r)$ is of order $M^2$ times larger than the $l$-dependent term according to the scaling law. Around the origin the potential approaches a negative value 
\begin{eqnarray}
V_l(0)\approx -\frac{b_2}{a_2}\left\{\begin{array}{l}
b_4/3\,,\quad\textrm{generic CQG}\\
\sqrt{a_2m_2^2}\,,\,\,\textrm{$\beta=0$ CQG}\end{array}\right. \,.
\end{eqnarray}
As seen in Fig.~\ref{fig:Vl22Schd solution}(a), this value turns out to be relatively small compared to the peak value of the potential at the light ring even for $l=1$.

From Fig.~\ref{fig:Vl22Schd solution}(b) we see that the significantly negative part of the potential $M^2V_l$ has a shape (depth and width) that becomes independent of $M$ at large $M$. That is it falls well within the scaling region. A negative potential leads one to wonder whether there is an eigenmode with negative $\omega^2$ which could indicate an instability. Because of the boundary condition $\Psi(0)=0$ such an eigenmode is not guaranteed to exist. Its absence requires that the negative part of the potential be sufficiently small in terms of its width and depth. For example if $V(r)$ is $-v$ for $r<a$ and zero for $r>a$ then $va^2\lesssim2.47$ is required. From our numerical solutions in both $\beta=0$ and generic CQG we find that the negative potential is sufficiently small, although curiously not by a wide margin.

\subsection{A sketch of the rotating 2-2-hole}
\label{sec:rotation}

It is a general result that a stationary, axisymmetric metric describing rotation and with a horizon must have an ergoregion, a region where $g_{tt}$ only has changed sign, that exists outside the horizon. The Kerr metric is the prime example. But there is no such requirement for an ergoregion when there is no horizon. From our experience with non-rotating case, we might expect that the metric of the rotating 2-2-hole should match the Kerr metric down to the radii where the higher curvature terms in the action suddenly become important. The question is at what radii does this occur. One possibility is that the strong gravity region extends out to the infinite red-shift surface of the Kerr metric. Then $g_{tt}$ becomes small but doesn't vanish, and then not only the horizon but also the exterior ergoregion of the Kerr metric is replaced by something else.\footnote{Note that for the Kerr metric with rapid rotation the light ring falls into the ergoregion. A modification of the latter by the strong gravity may have impact on the image of the object.} This is similar to a non-rotating 2-2-hole where strong gravity extends out to where $g_{tt}$ would vanish in the Schd metric. The difference in the rotating case is that $g_{rr}$ is not becoming as large as $1/g_{tt}$ at the radii where the curvatures are becoming large.

Let us consider some possible approximations to the interior of a rotating 2-2-hole. The following stationary and axisymmetric metric displays a rotation parameterized by a function $\omega(r)$.
\begin{align}
ds^2=-B(r)dt^2+A(r)dr^2+r^2d\theta^2+r^2\sin^2\theta[d\phi-\omega(r)dt]^2
\label{e6}\end{align}
An interesting feature of this metric is that the vacuum field equations are independent of $\omega(r)$ when it is a constant, $\omega(r)=\omega_0$. Here $\omega_0$ could be set equal to the angular velocity of rotating shell of matter and $A(r)$ and $B(r)$  in the interior would be the same as for the non-rotating 2-2-hole. Outside the 2-2-hole $\omega(r)$ should have a $1/r^3$ behavior to match the asymptotic Kerr metric. Then there would have to be a transition region between the interior and the exterior regions where a nontrivial $\theta$ dependence would have to enter the metric.

For this metric in the interior, $g_{tt}=-B(r)+\sin^2\theta \omega_0^2 r^2$ where $\omega_0 M\lesssim1$. In the interior $B(r)$ quickly falls to values where $B(r)/r^2\sim1/M^4$, and so for any appreciable rotation $g_{tt}$ will change sign and then remain positive down to $r=0$. Thus in this picture most of the interior of a rotating 2-2-hole is an ergoregion, except for a tiny cone around the poles ($\sin\theta\lesssim1/M$).

But we may also ask whether $\omega(r)$ could vary in the interior, and perhaps instead fall to zero at $r=0$, in such a way as to avoid an ergoregion. The $t$-$\phi$ field equation has only odd powers in $\omega(r)$ while the other field equations have even powers. We can treat $\omega(r)$ as small in the interior and study the $t$-$\phi$ equation at linear order in $\omega(r)$. We may use our series expansions of $A(r)$ and $B(r)$ to find a series expansion solution for $\omega(r)$. As we shall see a series expansion solution that starts with the highest possible power of $r$ is of interest, which turns out to be of the form
\begin{align}
\omega(r)=w\,(r^3 +b_4 r^5 +{\cal O}(r^7))
.\label{e1}\end{align}
In order for $g_{tt}$ not to change sign, $\omega(r)^2r^2$ must not exceed $B(r)$, which implies that $w\sim1/M^5$ or smaller.

We can now insert this series expansion for $\omega(r)$ into the other field equations. We find that the series expansions of the new terms in these equations, the terms quadratic in $\omega(r)$, start with the same power of $r$ as the expansion of the original terms. (If the leading power in $\omega(r)$ had been less than three then it would have led to corrections with a lower power of $r$, thus ruining the original $(2,2)$ family classification.) Most importantly the new terms are subdominant in powers of $1/M$ as compared to the original terms. Additional $\theta$ dependence shows up in these small corrections. Again there would have to be a transition region that matches such an interior solution to the exterior solution. From this discussion it seems possible for a rotating 2-2-hole to have no ergoregion and for the interior $A(r)$ and $B(r)$ functions to be little changed from the non-rotating case.

We now turn to the geodesics in a rotating 2-2-hole for the two possibilities of the interior region that we have considered. For the geodesics confined to the equatorial plane, the geodesic equation for a general $\omega(r)$ can be reduced to the following,
\begin{align}
\frac{d\phi}{d\zeta}&=\frac{L}{r^2}+\omega(r)\frac{E-\omega(r) L}{B(r)}\label{e3},\\
\frac{dt}{d\zeta}&=\frac{E-\omega(r) L}{B(r)}\label{e4},\\
\left(\frac{dr}{d\zeta}\right)^2&=\frac{1}{A(r)}\left(\frac{(E-\omega(r) L)^2}{B(r)}-\frac{L^2}{r^2}-\vartheta\right).\label{e2}
\end{align}
Since we can require that coordinate time moves forward for a particle moving along a geodesic we have $dt/d\zeta>0$. The angular velocity in coordinate time is
\begin{align}
\frac{d\phi}{dt}=\frac{L}{E-\omega(r) L}\frac{B(r)}{r^2}+\omega(r).
\label{e5}\end{align}

We first consider the case of no ergoregion where $\omega(r)$ is of the form of (\ref{e1}). From $dt/d\zeta>0$ and the vanishing of $\omega(r)$ at the origin we see that $E>0$. Also $E\gtrsim|\omega(r)L|$ since $\omega(r)^2<B(r)/r^2$ for no ergoregion and $E^2/L^2\approx B(r)/r^2$ when $r$ is the turning point of a bound interior orbit. The positivity of (\ref{e2}) further constrains $E$. In (\ref{e5}) $\omega(r)$ shows up as a simple frame dragging effect (second term) and also as a distortion of the original term. The two terms can be of the same order of magnitude and of equal or opposite sign. Depending on the sign of the $L\omega(r)$ term in (\ref{e2}), the radius of the turning point in the orbit can also increase or decrease. Thus the orbits are affected, but since $d\phi/dt$ still scales like $1/M^2$, the original near straight line motion still persists in the large $M$ limit.

The case where $\omega(r)=\omega_0$ and there is an interior ergoregion is quite different. Now $\omega_0$ scales like $1/M$ and so $|E|\ll|\omega_0 L|$. Then $\omega_0 L<0$ to have $dt/d\zeta>0$ while now the energy $E$ can have either sign. With large $\omega_0$ in (\ref{e5}) the frame dragging effect completely dominates. Thus the interior orbits can now significantly depart from near straight line motion.  But there are still no interior circular orbits. Their absence for the non-rotating 2-2-hole is due to the form of $B(r)$, and the introduction of the $\omega_0$ constant does not change this because it effectively just produces a shift in the constant $E$ in (\ref{e2}).

For ultra-compact stars the interior stable light rings are associated with resonant negative energy modes in wave equations when there is rotation. This leads to issues with the ergoregion instability that renders some ultra-compact stars too short-lived \cite{Friedman, Comins, Cardoso:2014sna, Moschidis:2016zjy}. For a rotating 2-2-hole with an interior ergoregion we have only found non circular orbits with negative energy. The relation that these have to any ergoregion instability remains to be studied. But if there exists a rotating 2-2-hole solution with no ergoregion anywhere, as our discussion has hinted, then it could be the preferred stable configuration.

\begin{acknowledgments}

This research is supported in part by the Natural Sciences and Engineering Research
Council of Canada. We are grateful for useful discussions with Vitor Cardoso, Akihiro Ishibashi and Roman Koniuk.

\end{acknowledgments}	

\appendix
\section{Series expansions}
\label{app:SeriesExp}

\subsection{Generic CQG}

We list explicit forms of the series expansion for generic CQG with the action parameters $\alpha\neq0$, $\beta\neq0$.  The expressions can be simplified with $m_2^2=\Mp^2/2\alpha$ and $m_0^2=\Mp^2/6\beta$. The $(0,0)$ family is characterized by two free parameters $(a_2, b_2)$,
\begin{eqnarray}
A(r)&=&1+a_2r^2+\frac{r^4}{30} \Big[3 a_2^2 \left(10 +m_2^2/m_0^2 \right)+a_2\left(2 m_0^2+m_2^2 -6   b_2 \right)\nonumber\\
&&-b_2 \left(2(m_0^2 -m_2^2 )+3  b_2 \left(2  + m_2^2/m_0^2  \right)\right)\Big]+O(r^6),\nonumber\\
\frac{B(r)}{b_0}&=&1+b_2r^2+\frac{r^4 }{60  }\Big[3 a_2^2 m_2^2/m_0^2+a_2 \left(m_2^2-m_0^2+18  b_2 \right)\nonumber\\
&&+b_2 \left(m_0^2 +2 m_2^2 +3b_2 (6 - m_2^2/m_0^2 )\right)\Big]+O(r^6)\,.
\end{eqnarray}
The series expansion starts to be sensitive to the action at $\mathcal{O}(r^4)$. 
The (2,2) family is characterized by five free parameters $(a_2, a_5, b_3, b_4, b_5)$,
\begin{eqnarray}
A(r)&=&a_2 r^2+a_2 r^3-\frac{a_2 }{6} \left(2 a_2 -8 b_4 +b_3^2\right)r^4+a_5 r^5+\mathcal{O}(r^6),\nonumber\\
\frac{B(r)}{b_2}&=&r^2+b_3 r^3+b_4 r^4+b_5 r^5+\mathcal{O}(r^6)\,.
\end{eqnarray}
Here the dependence on the action is delayed to $\mathcal{O}(r^6)$. The $(2,2)_E$ family can be derived by simply taking $a_{2n+1}=0$, which then leaves two free parameters $(a_2, b_4)$,
\begin{eqnarray}
\frac{A(r)}{a_2}&=&r^2-\frac{1}{3} \left( a_2 -4 b_4\right)r^4-\frac{1}{36 } \Big[a_2^2m_0^2/m_2^2+3 a_2 \left(10  b_4+3m_0^2 \right)\nonumber\\
&&-b_4^2 \left(54 + m_0^2/m_2^2\right)\Big] r^6+\mathcal{O}(r^8),\nonumber\\
\frac{B(r)}{b_2}&=&r^2+b_4 r^4-\frac{1}{18} \left(2 a_2 b_4+a_2^2-15 b_4^2\right) r^6+\mathcal{O}(r^8)\,.
\end{eqnarray}
Ar large $M$ we need only keep the leading order terms in the $1/M$ expansion. 
\begin{eqnarray}
\frac{A(r)}{a_2}&=&r^2+\frac{4}{3}b_4 r^4-\frac{1}{36 } \left(9 a_2 m_0^2 -b_4^2 \left(54 + m_0^2/m_2^2 \right)\right) r^6
-\frac{19b_4}{810 } \big(27 a_2 m_0^2 \nonumber\\
&&-b_4^2 \left(70 + 3m_0^2/m_2^2 \right)\big) r^8+O(r^{10}),\nonumber\\
\frac{B(r)}{b_2}&=&r^2+b_4 r^4+\frac{5}{6}  b_4^2 r^6
-\frac{b_4}{180}\left(9a_2m_0^2-b_4^2\left(125+m_0^2/m_2^2\right)\right)r^8+O(r^{10})\,.
\end{eqnarray}
We see that the dependence on $\Mp$ still survives in this limit. If one considered a vanishing $\Mp$ then one would be left with a dependence on the combinations $b_4 r^2$ and $\alpha/\beta$.

\subsection{$\beta=0$ CQG}

With $m_2^2=\Mp^2/2\alpha$, the $(0,0)$ family is characterized by one free parameter $b_2$,
\begin{eqnarray}
A(r)&=&1+b_2r^2+\frac{b_2}{10 }\left(6  b_2+m_2^2 \right)r^4+\frac{b_2 }{280}\left(80 b_2^2+50  b_2 m_2^2 +m_2^4\right)r^6+O(r^8),\nonumber\\
\frac{B(r)}{b_0}&=&1+b_2r^2+\frac{b_2}{20} \left(12 b_2+m_2^2 \right)r^4+\frac{b_2 }{840}\left(240 b_2^2+72  b_2 m_2^2 + m_2^4\right)r^6+O(r^8)\,.
\end{eqnarray}
The (2,2) family is characterized by three free parameters $(a_2, b_3, b_4)$,
\begin{eqnarray}
\frac{A(r)}{a_2}&=&r^2+b_3 r^3-\frac{1}{6}\left(2 a_2-8 b_4+b_3^2\right)r^4
+\frac{1}{18  b_3}  \Big(10 a_2^2+a_2 \left(11 b_3^2+90 m_2^2 \right)\nonumber\\
&&+ 12 b_3^4-25 b_4 b_3^2-10 b_4^2\Big) r^5+O(r^6),\nonumber\\
\frac{B(r)}{b_2}&=&r^2+b_3r^3+b_4r^4 
-\frac{1}{18 b_3}\left(6 a_2^2+a_2 \left( b_3^2+54 m_2^2\right)+8 b_3^4-19 b_4 b_3^2-6 b_4^2 \right)r^5 \nonumber\\
&&+O(r^8)\,.
\end{eqnarray}
The series expansion starts to be sensitive to the action at $\mathcal{O}(r^5)$. As $b_3$ appears in the denominator, we cannot derive the $(2,2)_E$ family by directly switching off the odd terms. Instead the $(2,2)_E$ family is characterized by only one free parameter $a_2$,
\begin{eqnarray}
\frac{A(r)}{a_2}&=&r^2-\frac{r^4 }{3}\left[a_2-4 \sqrt{a_2 \left(a_2+9 m_2^2\right)}\right]+\frac{a_2}{6}\left[9 a_2+81 m_2^2-5 \sqrt{a_2 \left(a_2+9 m_2^2 \right)}\right]r^6 +O(r^8),\nonumber\\
\frac{B(r)}{b_2}&=&r^2+r^4 \sqrt{a_2 \left( a_2+9m_2^2 \right)}+\frac{a_2}{18 } \left[-2 \sqrt{a_2 \left(a_2+9 m_2^2 \right)}+14 a_2+135 m_2^2 \right]r^6+O(r^8)\,.
\end{eqnarray}
The series expansion in the large $M$ limit is
\begin{eqnarray}
\frac{A(r)}{a_2}&=&r^2\left[1+4   \left(a_2 m_2^2 r^4\right)^{1/2}  +\frac{27}{2 } a_2 m_2^2 r^4 +\frac{133}{3} \left(a_2 m_2^2 r^4\right)^{3/2}+O(r^{8}) \right],\nonumber\\
\frac{B(r)}{b_2}&=&r^2\left[1+3   \left(a_2 m_2^2 r^4\right)^{1/2}  +\frac{15}{2 } a_2 m_2^2 r^4 +\frac{75}{4} \left(a_2 m_2^2 r^4\right)^{3/2}+O(r^{8}) \right]
.\end{eqnarray}
The essential $\Mp$ dependence here is related to the conformal invariance of the theory with vanishing $\Mp$.



\begin{thebibliography}{99}

\bibitem{Mathur:2005zp} 
  S.~D.~Mathur,
  Fortsch.\ Phys.\  {\bf 53}, 793 (2005)
  [hep-th/0502050].

\bibitem{Almheiri:2012rt} 
  A.~Almheiri, D.~Marolf, J.~Polchinski and J.~Sully,
  JHEP {\bf 1302}, 062 (2013)
  [arXiv:1207.3123 [hep-th]].

\bibitem{Giddings:2014ova} 
  S.~B.~Giddings,
  Phys.\ Rev.\ D {\bf 90}, no. 12, 124033 (2014)
  [arXiv:1406.7001 [hep-th]].

\bibitem{Stelle:1976gc}
  K.~S.~Stelle,
  Phys.\ Rev.\ D {\bf 16}, 953 (1977).

\bibitem{Voronov:1984kq}
  B.~L.~Voronov and I.~V.~Tyutin,
  Yad.\ Fiz.\  {\bf 39}, 998 (1984).

\bibitem{Fradkin:1981iu}
  E.~S.~Fradkin and A.~A.~Tseytlin,
  Nucl.\ Phys.\ B {\bf 201}, 469 (1982).

\bibitem{Avramidi:1985ki}
  I.~G.~Avramidi and A.~O.~Barvinsky,
  Phys.\ Lett.\ B {\bf 159}, 269 (1985).


\bibitem{Holdom:2015kbf} 
  B.~Holdom and J.~Ren,
  Phys.\ Rev.\ D {\bf 93}, no. 12, 124030 (2016)
  [arXiv:1512.05305 [hep-th]]; Int.\ J.\ Mod.\ Phys.\ D {\bf 25}, no. 12, 1643004 (2016)
  [arXiv:1605.05006 [hep-th]].

\bibitem{Donoghue:2016vck} 
  J.~F.~Donoghue,
  arXiv:1609.03523 [hep-th]; arXiv:1609.03524 [hep-th].


\bibitem{Stelle:1977ry} 
  K.~S.~Stelle,
  Gen.\ Rel.\ Grav.\  {\bf 9}, 353 (1978).

\bibitem{Holdom:2002xy} 
  B.~Holdom,
  Phys.\ Rev.\ D {\bf 66}, 084010 (2002)
  [hep-th/0206219].

\bibitem{BHM1} 
  P.~O.~Mazur and E.~Mottola,
  gr-qc/0109035.

\bibitem{BHM2} 
  D.~J.~Kaup,
  Phys.\ Rev.\  {\bf 172}, 1331 (1968).

\bibitem{BHM3} 
  R.~Ruffini and S.~Bonazzola,
  Phys.\ Rev.\  {\bf 187}, 1767 (1969).

\bibitem{BHM4} 
  T.~Damour and S.~N.~Solodukhin,
  Phys.\ Rev.\ D {\bf 76}, 024016 (2007)
  [arXiv:0704.2667 [gr-qc]].  
 
 \bibitem{Visser:2003ge} 
  M.~Visser and D.~L.~Wiltshire,
  Class.\ Quant.\ Grav.\  {\bf 21}, 1135 (2004)
  doi:10.1088/0264-9381/21/4/027
  [gr-qc/0310107].

\bibitem{Visser:2009pw} 
  M.~Visser, C.~Barcelo, S.~Liberati and S.~Sonego,
  PoS BHGRS {\bf }, 010 (2008)
  [arXiv:0902.0346 [gr-qc]].
  
 \bibitem{Wald:1980jn} 
  R.~M.~Wald,
  J.\ Math.\ Phys.\  {\bf 21}, 2802 (1980).

\bibitem{Abbott:2016blz} 
  B.~P.~Abbott {\it et al.} [LIGO Scientific and Virgo Collaborations],
  Phys.\ Rev.\ Lett.\  {\bf 116}, no. 6, 061102 (2016)
  [arXiv:1602.03837 [gr-qc]];
  Phys.\ Rev.\ X {\bf 6}, no. 4, 041015 (2016)
  [arXiv:1606.04856 [gr-qc]].
  
\bibitem{Cardoso:2016rao} 
  V.~Cardoso, E.~Franzin and P.~Pani,
  Phys.\ Rev.\ Lett.\  {\bf 116}, no. 17, 171101 (2016)
  [arXiv:1602.07309 [gr-qc]].

\bibitem{Cardoso:2016oxy} 
  V.~Cardoso, S.~Hopper, C.~F.~B.~Macedo, C.~Palenzuela and P.~Pani,
  arXiv:1608.08637 [gr-qc].

\bibitem{Abedi:2016hgu} 
  J.~Abedi, H.~Dykaar and N.~Afshordi,
  arXiv:1612.00266 [gr-qc].

  

\bibitem{Lu:2015psa} 
  H.~Lu, A.~Perkins, C.~N.~Pope and K.~S.~Stelle,
  Phys.\ Rev.\ D {\bf 92}, no. 12, 124019 (2015)
  [arXiv:1508.00010 [hep-th]].


\bibitem{Lu:2015cqa} 
  H.~Lu, A.~Perkins, C.~N.~Pope and K.~S.~Stelle,
  Phys.\ Rev.\ Lett.\  {\bf 114}, no. 17, 171601 (2015)
  [arXiv:1502.01028 [hep-th]].
  
 \bibitem{Geroch:1987qn} 
  R.~P.~Geroch and J.~H.~Traschen,
  Phys.\ Rev.\ D {\bf 36}, 1017 (1987)
  [Conf.\ Proc.\ C {\bf 861214}, 138 (1986)].
  
  \bibitem{Israel:1966rt} 
  W.~Israel,
  Nuovo Cim.\ B {\bf 44S10}, 1 (1966)
  [Nuovo Cim.\ B {\bf 44}, 1 (1966)]
  Erratum: [Nuovo Cim.\ B {\bf 48}, 463 (1967)].

\bibitem{Cardoso:2014sna} 
  V.~Cardoso, L.~C.~B.~Crispino, C.~F.~B.~Macedo, H.~Okawa and P.~Pani,
  Phys.\ Rev.\ D {\bf 90}, no. 4, 044069 (2014)
  [arXiv:1406.5510 [gr-qc]].

\bibitem{Harada:2014vka} 
  See a review, T.~Harada and M.~Kimura,
  Class.\ Quant.\ Grav.\  {\bf 31}, 243001 (2014)
  [arXiv:1409.7502 [gr-qc]].

\bibitem{Patil:2012fu} 
  M.~Patil and P.~S.~Joshi,
  Phys.\ Rev.\ D {\bf 86}, 044040 (2012)
  [arXiv:1203.1803 [gr-qc]].

\bibitem{Abramowicz:2002vt} 
  M.~A.~Abramowicz, W.~Kluzniak and J.~P.~Lasota,
  Astron.\ Astrophys.\  {\bf 396}, L31 (2002)
  [astro-ph/0207270].

\bibitem{Broderick:2009ph} 
A.~E.~Broderick, A.~Loeb and R.~Narayan, 
Astrophys.\ J.\  {\bf 701}, 1357 (2009)
[arXiv:0903.1105 [astro-ph.HE]].


\bibitem{Horowitz:1995gi} 
  G.~T.~Horowitz and D.~Marolf,
  Phys.\ Rev.\ D {\bf 52}, 5670 (1995)
  [gr-qc/9504028].
  
  
\bibitem{Ishibashi:1999vw} 
  A.~Ishibashi and A.~Hosoya,
  Phys.\ Rev.\ D {\bf 60}, 104028 (1999)
  [gr-qc/9907009].  
  
\bibitem{Ishibashi:2003jd}
A.~Ishibashi and R.~M.~Wald,
 Class.\ Quant.\ Grav.\  {\bf 20}, 3815 (2003)
 [gr-qc/0305012].

\bibitem{Chirenti:2012fr} 
  C.~Chirenti, A.~Saa and J.~Skakala,
  Phys.\ Rev.\ D {\bf 86}, 124008 (2012)
  [arXiv:1206.0037 [gr-qc]].  


\bibitem{Horowitz:1995ta} 
  G.~T.~Horowitz and R.~C.~Myers,
  Gen.\ Rel.\ Grav.\  {\bf 27}, 915 (1995)
  [gr-qc/9503062].








\bibitem{Friedman} 
 J. L. Friedman. 
 Communications in Mathematical Physics, 63(3):243, 1978.

\bibitem{Comins} 
 N. Comins and B. F. Schutz, Proc. R. Soc. Lond. A 364, 211 (1978).
  

\bibitem{Moschidis:2016zjy} 
  G.~Moschidis,
  arXiv:1608.02035 [math.AP].


\bibitem{tHooft:1984kcu} 
  G.~'t Hooft,
  Nucl.\ Phys.\ B {\bf 256}, 727 (1985).
  
\bibitem{Mukohyama:1998rf} 
  S.~Mukohyama and W.~Israel,
  Phys.\ Rev.\ D {\bf 58}, 104005 (1998)
  [gr-qc/9806012].
  
\end{thebibliography}
\end{document}